\documentstyle[12pt,epsfig]{report}

\makeatletter
\def\dash{-}
\def\@citex[#1]#2{%
\if@filesw\immediate\write\@auxout{\string\citation{#2}}\fi
  \def\@citea{}\@cite{\@for\@citeb:=#2\do
     {\ifx\dash\@citeb{--}\def\@citea{}\else
      \@citea\def\@citea{,\penalty\@m\ }\@ifundefined
       {b@\@citeb}{{\bf ?}\@warning
       {Citation `\@citeb' on page \thepage \space undefined}}\fi
\hbox{\csname b@\@citeb\endcsname}}}{#1}}
\newbox\tempboxa
\newdimen\captionboxsubcount 
\def\capsize#1{\captionboxsubcount=#1pt}
\newdimen\captionboxsub
\captionboxsub=\hsize \advance\captionboxsub by -\captionboxsubcount
\advance\captionboxsub by -\captionboxsubcount
\long\def\@makecaption#1#2{
 \setbox\@tempboxa\hbox{#1: #2}
 \ifdim \wd\@tempboxa >\captionboxsub 
\rightskip=\captionboxsubcount \leftskip=\captionboxsubcount #1: #2 
\else \hbox to\hsize{\hfil\box\@tempboxa\hfil} 
 \fi}
\makeatother
\capsize{30}

\def\thebibliography#1{\leftline{\Large \it References}\list
  {[\arabic{enumi}]}{\settowidth\labelwidth{[#1]}\leftmargin\labelwidth
    \advance\leftmargin\labelsep
    \usecounter{enumi}}
    \def\newblock{\hskip .11em plus .33em minus .07em}
    \sloppy\clubpenalty4000\widowpenalty4000}

\newcommand{\be}{\begin{eqnarray}}
\newcommand{\ee}{\end{eqnarray}}
\newcommand{\bd}{\begin{displaymath}}
\newcommand{\ed}{\end{displaymath}}

\newcommand{\eal}{\left(\epsilon-\frac{\alpha}{2}\omega\right)}
\newcommand{\rfp}{\sqrt{4\pi}}
\newcommand{\freq}{{\rm e}^{i\epsilon t}}

\begin{document}

\rightline{June, 1997}

\bigskip

\begin{center}
\Large\bf
Hyperfine Splitting of Low--Lying Heavy Baryons
\end{center}

\bigskip

\begin{center}
\large
Masayasu {\sc Harada}$^{(a)}$\footnote{
Electronic address : {\tt mharada@npac.syr.edu}},
Asif {\sc Qamar}$^{(a)}$\footnote{
Electronic address : {\tt qamar@suhep.phy.syr.edu}},
Francesco {\sc Sannino}$^{(a,b)}$\footnote{
Electronic address : {\tt sannino@npac.syr.edu}},\\
Joseph {\sc Schechter}$^{(a)}$\footnote{
Electronic address : {\tt schechter@suhep.phy.syr.edu}}
and
Herbert {\sc Weigel}$^{(c)}$\footnote{
Electronic address : 
{\tt weigel@sunelc1.tphys.physik.uni-tuebingen.de}}
\end{center}

\bigskip

\begin{flushleft}
\it 
\qquad$^{(a)}$Department of Physics, Syracuse University, 
Syracuse, NY 13244-1130, USA
\\
\qquad$^{(b)}$%
Dipartimento di Scienze Fisiche \& Istituto
Nazionale di Fisica Nucleare \\
\qquad\qquad Mostra D'Oltremare Pad. 19,  80125 Napoli, Italy
\\
\qquad$^{(c)}$%
Institute for Theoretical Physics, 
T\"ubingen University \\
\qquad\qquad
Auf der Morgenstelle 14, D-72076 T\"ubingen, Germany
\end{flushleft}

\bigskip

\centerline{\Large \bf Abstract}
\bigskip

We calculate the next--to--leading 
order contribution to the masses of the heavy baryons in the bound 
state approach for baryons containing a heavy quark. These $1/N_C$ 
corrections arise when states of good spin and isospin are generated 
from the background soliton of the light meson fields. Our study
is motivated by the previously established result that light vector 
meson fields are required for this soliton in order to reasonably 
describe the spectrum of both the light and the heavy baryons.
We note that the inclusion of light vector mesons significantly 
improves the agreement of the predicted hyperfine splitting with 
experiment.  A number of aspects of this somewhat complicated 
calculation are discussed in detail.

\bigskip
\begin{flushleft}
\small
PACS numbers: 12.39.Dc, 12.39.Fe, 12.39.Hg \\
Keywords: \parbox[t]{12cm}{Heavy spin symmetry, $1/M$ expansion, 
Skyrmions, heavy quark solitons, collective quantization, 
hyperfine splitting}
\end{flushleft}

\newpage

\stepcounter{chapter}
\leftline{\Large \it 1. Introduction}
\smallskip

The development of the heavy quark or Isgur-Wise symmetry~\cite{Ei81}
has stimulated a great deal of interest in 
studying~\cite{Je92}\nocite{Rho93,Gu93,Mo94,Oh94}--\cite{Oh94a}
properties of heavy baryons (i.e., those with the quark structure 
$qqQ$) in the bound state approach. In this picture the heavy 
baryon is treated as a heavy spin multiplet of mesons (with 
structure $Q\bar{q}$) bound in the field of the nucleon ($qqq$) which 
itself emerges as a soliton configuration of light meson fields. This 
treatment is suggested by the $1/N_{\rm C}$ expansion~\cite{Wi79} 
of QCD.  Recent reviews of the soliton approach to the light 
baryons are given in refs~\cite{Ho86}\nocite{Me88,Al96}--\cite{We96} 
while the bound state treatment of the ``light'' hyperons is discussed 
in refs~\cite{Ca85,Bl88}.

A compelling feature of this approach is that it permits, in principle,
an exact expansion of the heavy baryon properties in simultaneous 
powers of $1/M$, $1/N_{\rm C}$ and, since it is based on a chiral 
Lagrangian, number of derivatives acting on the light components 
of the heavy system. In practice there are obstacles related to the 
large number of unknown parameters which must be introduced.
Rather than treating the light soliton in a model with many 
derivatives of the light pseudoscalar fields it turns out to be 
much more efficient to use the light vector mesons. Based on a 
model~\cite{Sch93} of the light vector interactions with the 
heavy multiplet, the leading order (in the $1/N_{\rm C}$ and $1/M$ 
expansions) heavy baryon mass splittings have been discussed 
\cite{Sch96}, obtaining satisfactory agreement with experiment. 
Actually the need for light vector mesons is not surprising since, in 
the soliton approach, they are necessary to explain, for example, 
the neutron--proton mass difference~\cite{Ja89} and the nucleon
axial singlet matrix element~\cite{Jo90}.

In the present paper we focus our attention on the hyperfine splitting,
which is of subleading order both in $1/M$ and $1/N_{\rm C}$.
This is a more complicated calculation and also involves using a 
cranking procedure~\cite{Ad83} to obtain physical states which
carry good spin and isospin quantum numbers. The first calculation 
of the heavy baryon hyperfine splitting in the perturbative
bound state framework was carried out by Jenkins and Manohar~\cite{Je92}
who got the formula
\begin{equation}
m(\Sigma_Q^{\ast}) - m(\Sigma_Q) =
\frac{\left( m(\Delta) - m(N) \right) \left(M^{\ast} - M\right)}{
4d\,F'(0)}
\ ,
\label{eq:1.1}
\end{equation}
where $M^{\ast}-M$ is the heavy vector--heavy pseudoscalar mass 
difference, $d$ is the light pseudoscalar--heavy meson coupling 
constant and $F'(0)$ is the slope of the Skyrme ``profile function''
at the origin.  This formula is obtained
(see also section 5) by using the leading order 
in number of derivatives (zero) and leading order in $1/M$ heavy 
spin violation term.  Therefore it is expected to provide the 
dominant contribution. Unfortunately, on evaluation, it is found 
to provide only a small portion of the experimental 
$\Sigma_c^{\ast}$--$\Sigma_c$ masss difference. This naturally 
suggests the need for including additional higher order in derivative 
heavy spin violation terms. However, there are many possible terms 
with unknown coefficients so that the systematic perturbative approach 
is not very predictive.

To overcome this problem we employ a relativistic Lagrangian 
model~\cite{Sch93} which uses ordinary heavy pseudoscalar and 
vector fields rather than the heavy ``fluctuation'' field 
multiplet~\cite{Ei81}. This model reduces to the heavy multiplet 
approach in leading order and does not contain any new parameters.
In a recent note~\cite{Ha97} we showed that such a model 
(considered, for simplicity, to contain only light pseudoscalars;
{\it i.e.}, the light part is the original Skyrme model~\cite{Sk61})
yielded a ``hidden'' heavy spin violation which is not manifest 
from the form of the Lagrangian itself. This hidden part involves 
two derivatives and is actually more important numerically than the 
zero derivative ``manifest'' piece which leads to eq~(\ref{eq:1.1}).
However this new result is still not sufficient to bring the 
predicted $\Sigma_c^{\ast}$--$\Sigma_c$ mass difference into 
agreement with experiment. The prediction for this 
difference is actually 
correlated to those for 
 $\Sigma_c$--$\Lambda_c$ and $\Delta$--$N$, the 
  $\Delta$ - nucleon  mass difference by \cite{Ca85}:
\be
m\left(\Sigma_c^{\ast}\right)-m\left(\Sigma_c\right)
=m\left(\Delta\right)-m\left(N\right)-\frac{3}{2}\left[
m\left(\Sigma_c\right)-m\left(\Lambda\right)\right] \ .
\label{ck63}
\ee
This formula depends only on the collective quantization procedure 
being used rather than the detailed structure of the model.
If $m\left(\Sigma_c\right)-m\left(\Lambda\right)$ and 
 $m\left(\Delta\right)-m\left(N\right)$ are taken to agree with 
experiment, eq~(\ref{ck63}) predicts $41$~MeV rather than the 
experimental value of $66$~MeV. This means that it is impossible 
to exactly predict, in models of the present type, all three 
mass differences which appear in eq~(\ref{ck63}). The goodness 
of the overall fit must be judged by comparing all three 
quantities with experiment. Our focus, of course, is the left hand side 
of eq~(\ref{ck63}) which is of order $1/M$ while the right hand 
side involves the difference of two order $M^0$ quantities. 
A similar calculation in the model with 
only light pseudoscalars was carried out by Oh and Park~\cite{Oh96}.
However, they did not make a $1/M$ expansion in order to reveal the 
hidden violation terms. They also introduced a one--derivative 
``manifest'' heavy spin violation term with a new relatively large 
unknown constant in order to improve the agreement with experiment.

In the present paper we show that it is not necessary to introduce
any new violation terms to agree with experiment if a chiral 
Lagrangian including light vectors is employed. Typical results are 
summarized, compared with experiment and compared with the Skyrme 
model for the light sector in Table~\ref{tab:1.1}.
\begin{table}[htbp]
\caption[]{
Typical results for the present model (including light vectors)
compared with model with light pseudoscalars only (``Skyrme'' column)
and compared with experiment. No ``manifest'' heavy spin violation 
effects other than $M^{\ast}\neq M$ have been included. The column 
``present model + CM'' simply takes into account recoil corrections 
by replacing the heavy meson mass by the reduced mass. $\Lambda'_c$ 
denotes a negative parity, spin $1/2$ state. The quantity $\alpha$ 
in eqs~(\ref{covderp}) and (\ref{covderq}) was taken to be zero.
All masses in MeV.}
\label{tab:1.1}
\begin{center}
\begin{tabular}{|c||c|c|c|c|}
\hline
mass difference & expt. & present model
& present model + CM & Skyrme \\
\hline\hline
$\Lambda_c-N$ & 1345 & 1257 & 1356 & 1553 \\
$\Lambda_b-\Lambda_c$ & 3356$\pm$50 & 3164 & 3285 & 3215 \\
$\Lambda'_c-\Lambda_c$ & 308 & 249 & 342 & 208 \\
$\Sigma_c-\Lambda_c$ & 168 & 172 & 158 & 185 \\
$\Sigma_c^{\ast}- \Sigma_c$ & 66 & 42 & 63 & 16\\
\hline
\end{tabular}
\end{center}
\end{table}
A much more detailed discussion is given later in the text.
We  notice from the last row, that the model with light 
vectors gives a very satisfactory account of the 
$\Sigma_c^{\ast}$--$\Sigma_c$ hyperfine splitting in 
contrast to the model without light vectors. There are also 
noticeable effects when the use of the heavy meson reduced 
mass is taken as a simple approximation for kinematical corrections.
Similarly, the first four rows of Table~\ref{tab:1.1} show that 
the other predictions of the model with light vectors agree well 
with experiment.  Note that (see section 4) the predictions for 
mass differences are considered more reliable than those for the 
masses themselves.

The present article is organized as follows. In section 2 we will 
review the classical, leading in $1/N_C$, part of this calculation. 
This discussion includes both the light and heavy meson pieces of the 
Lagrangian. The emergence of bound state solutions is also
explained in section 2. In section 3 we will describe the collective 
quantization in the framework of the cranking procedure for the 
bound states. Section 4 contains a detailed discussions of the 
numerical results. In section 5 we will discuss some new manifest 
contributions to the hyperfine splittings and the extension to 
different channels in the framework of the perturbation approach. 
We will conclude in section 6. The explicit expressions for the 
couplings between the bound heavy meson and the collective coordinates 
are listed in the appendices.

\bigskip
\stepcounter{chapter}
\leftline{\Large \it 2. The Model Lagrangian}
\smallskip

In this section we review the classical, {\it i.e.} leading 
order part in the $1/N_C$ expansion of the bound state 
description for the heavy baryons in the soliton picture.

\bigskip
\leftline{\large \it 2.1. Light Mesons}
\smallskip

For the sector of the model describing the light pseudoscalar
and vector mesons we adopt the chirally invariant Lagrangian
discussed in detail in the literature \cite{Ka84,Ja88}.
This Lagrangian can be decomposed into a regular parity part
\begin{eqnarray}
{\cal L}_S=f_\pi^2{\rm tr}\left[p_\mu p^\mu\right]
+\frac{m_\pi^2f_\pi^2}{2}{\rm tr}\left[U+U^{\dag}-2\right]
-\frac{1}{2}{\rm tr}\left[F_{\mu\nu}\left(\rho\right)
F^{\mu\nu}\left(\rho\right)\right]
+m_V^2{\rm tr}\left[R_\mu R^\mu\right]
\label{lagnorm}
\end{eqnarray}
and a part which contains the Levi-Civita tensor,
$\epsilon_{\mu\nu\alpha\beta}$. The action for the latter is most
conveniently displayed with the help of differential 
forms $p=p_\mu dx^\mu$, etc.
\begin{eqnarray}
\Gamma_{\rm an}&&=\frac{2N_c}{15\pi^2}\int Tr(p^5)\cr
&&\quad
+\int Tr\left[\frac{4i}{3}(\gamma_1+\frac{3}{2}\gamma_2)Rp^3
-\frac{g}{2} \gamma_2 F(\rho )(pR-Rp)
-2ig^2 (\gamma_2+2\gamma_3) R^3p\right].
\label{lagannorm}
\end{eqnarray}
In eqs (\ref{lagnorm}) and (\ref{lagannorm}) we have introduced the
abbreviations
\begin{eqnarray}
p_\mu=\frac{i}{2}\left(\xi\partial_\mu\xi^{\dag}-
\xi^{\dag}\partial_\mu\xi\right),\quad
v_\mu=\frac{i}{2}\left(\xi\partial_\mu\xi^{\dag}+
\xi^{\dag}\partial_\mu\xi\right) \quad {\rm and}\quad
R_\mu=\rho_\mu-\frac{1}{g}v_\mu \ .
\label{lightcurrents}
\end{eqnarray}
Here $\xi$ refers to a square root of the chiral field,
{\it i.e.} $U=\xi^2$. Furthermore
$F_{\mu\nu}\left(\rho\right)=\partial_\mu\rho_\nu
-\partial_\nu\rho_\mu-ig\left[\rho_\mu,\rho_\nu\right]$
denotes the field tensor associated with the vector mesons
$\rho$ and $\omega$, which are combined in  
$\rho_\mu=\left(\omega_\mu{\sl 1\!\! I}+\rho_\mu^a\tau^a\right)/2$
when the reduction to two light flavors is made.  The parameters 
$g,\gamma_1,$ etc. can be determined (or at least constrained) 
from the study of decays of the light vector mesons such as 
$\rho\rightarrow2\pi$ or $\omega\rightarrow3\pi$ \cite{Ja88}. 

The action for the light degrees of freedom 
($\int {\cal L}_S+\Gamma_{\rm an}$) contains static soliton
solutions. The appropriate {\it ans\"atze} are
\begin{eqnarray}
\xi(\mbox{\boldmath $r$})={\rm exp}\left(\frac{i}{2}
\hat{\mbox{\boldmath $r$}}\cdot\mbox{\boldmath $\tau$}F(r)\right),
\quad
\omega_0(\mbox{\boldmath $r$})=\frac{\omega(r)}{g}
\quad {\rm and}\quad
\rho_{i,a}(\mbox{\boldmath $r$})=
\frac{G(r)}{gr}\epsilon_{ija}\hat r_j
\label{lightan}
\end{eqnarray}
while all other field components vanish. The resulting non--linear 
Euler--Lagrange equations for the radial functions $F(r),\omega(r)$ 
and $G(r)$ are solved numerically subject to the boundary conditions
$F(0)=-\pi$, $\omega^\prime(0)=0$ and $G(0)=-2$ while all fields vanish
at radial infinity \cite{Ja88}. These boundary conditions are needed 
to obtain a consistent baryon number one configuration.

\bigskip
\leftline{\large \it 2.2. The Relativistic Model for the Heavy Mesons}
\smallskip

In this subsection we present the relativistic Lagrangian, which 
describes the coupling between the light and heavy mesons \cite{Sch93}
\be
{\cal L}_H&=&D_\mu P\left(D^\mu P\right)^{\dag}
-\frac{1}{2}Q_{\mu\nu}\left(Q^{\mu\nu}\right)^{\dag}
-M^2PP^{\dag}+M^{*2}Q_\mu Q^{\mu{\dag}}
\nonumber \\ &&
+2iMd\left(Pp_\mu Q^{\mu{\dag}}-Q_\mu p^\mu P^{\dag}\right)
-\frac{d}{2}\epsilon^{\alpha\beta\mu\nu}
\left[Q_{\nu\alpha}p_\mu Q_\beta^{\dag}+
Q_\beta p_\mu \left(Q_{\nu\alpha}\right)^{\dag}\right]
\label{lagheavy} \\ &&
-\frac{2\sqrt{2}icM}{m_V}\left\{
2Q_\mu F^{\mu\nu}\left(\rho\right)Q_\nu^{\dag}
-\frac{i}{M}\epsilon^{\alpha\beta\mu\nu}\left[
D_\beta PF_{\mu\nu}\left(\rho\right)Q_\alpha^{\dag}
+Q_\alpha F_{\mu\nu}\left(\rho\right)\left(D_\beta P\right)^{\dag}
\right]\right\}.
\nonumber
\ee
Here we have allowed the mass $M$ of the heavy pseudoscalar meson
$P$ to differ from the mass $M^*$ of the heavy vector meson $Q_{\mu}$.
Note that the heavy meson fields are conventionally defined as {\it row}
vectors in isospin space. The covariant derivative introduces the 
additional parameter $\alpha$:
\begin{eqnarray}
D_\mu P^{\dag}&=&\left(\partial_\mu-i\alpha g \rho_\mu
-i\left(1-\alpha\right)v_\mu\right)P^{\dag}
=\left(\partial_\mu-iv_\mu-ig\alpha R_\mu\right)P^{\dag}\ ,
\label{covderp} \\
D_\mu Q_\nu^{\dag}&=&
\left(\partial_\mu-iv_\mu-ig\alpha R_\mu\right)Q_\nu^{\dag} \ .
\label{covderq}
\end{eqnarray}
The covariant field tensor of the heavy vector meson is then 
defined as
\begin{eqnarray}
\left(Q_{\mu\nu}\right)^{\dag}=
D_\mu Q_\nu^{\dag}-D_\nu Q_\mu^{\dag}.
\label{heavyft}
\end{eqnarray}
The coupling constants $d,c$ and $\alpha$, which appear in the 
Lagrangian (\ref{lagheavy}), have still not been very accurately 
determined. In particular there is no direct experimental evidence 
for the value of $\alpha$, which would be unity if a possible 
definition of light vector meson dominance for the electromagnetic 
form factors of the heavy mesons were to be adopted \cite{Ja95}. We 
will later adjust $\alpha$ to the spectrum of the heavy baryons. The 
other parameters in (\ref{lagheavy}) will be taken~\cite{Ja95} to be:
\begin{eqnarray}
d&=&0.53\ , \quad c=1.60\ ;
\nonumber \\
M&=&1865{\rm MeV}\ , \quad M^*=2007\,{\rm MeV}\ , \qquad {\rm D-meson}\ ;
\nonumber \\
M&=&5279{\rm MeV}\ , \quad M^*=5325\,{\rm MeV}\ , \qquad {\rm B-meson}.
\label{heavypara}
\end{eqnarray}

It should be stressed that the assumption of infinitely large
masses for the heavy mesons has not been made in (\ref{lagheavy}).
However, a model Lagrangian which was only required to exhibit the 
Lorentz and chiral invariances would be more general than the 
relativistic Lagrangian (\ref{lagheavy}). The additional restrictions 
arise from the heavy quark transformation
\begin{eqnarray}
P^\prime = {\rm e}^{iM V\cdot x} P \ , \qquad
Q_\mu^\prime = {\rm e}^{iM^* V\cdot x} Q_\mu \ ,
\label{hqt}
\end{eqnarray}
where the four--velocity $V^\mu$ characterizes the reference frame 
of the heavy quark. The heavy pseudoscalar and vector meson fields 
may then be combined in the heavy multiplet
\begin{eqnarray}
H=\frac{1}{2}\left(1+\gamma_\mu V^\mu\right)
\left(i\gamma_5 P^\prime+\gamma^\nu Q^\prime_\nu\right)
\qquad {\rm and} \qquad {\bar H}=\gamma_0H^{\dag}\gamma_0\ .
\label{defh}
\end{eqnarray}
In the heavy quark limit, $M=M^*\to\infty$, the relativistic 
Lagrangian (\ref{lagheavy}) becomes
\begin{eqnarray}
\frac{1}{M}{\cal L}_H\to 
i V^\mu {\rm Tr}\left\{H D_\mu {\bar H}\right\}
- d {\rm Tr}\left\{H\gamma_\mu\gamma_5 p^\mu {\bar H}\right\}
- i \frac{\sqrt{2}c}{m_V}{\rm Tr}\left\{H\gamma_\mu\gamma_\nu 
F^{\mu\nu}(\rho){\bar H}\right\} + \ldots \ .
\label{laghl}
\ee
The ellipses indicate subleading pieces in $1/M$. Actually the 
coefficients of the various Lorentz and chirally invariant pieces 
in the relativistic Lagrangian (\ref{lagheavy}) have precisely been 
arranged to yield the spin--flavor symmetric model (\ref{laghl}) 
in the heavy quark limit \cite{Sch93}.

\bigskip
\leftline{\large \it 2.3. Bound States}
\smallskip

Here we briefly review the origin of bound states in the S-- and 
P--wave heavy meson channels. These orbital angular momentum quantum 
numbers refer to those of the pseudoscalar component ($P^{\dag}$) of 
the heavy meson multiplet ($P^{\dag},Q_\mu^{\dag}$).

For the P--wave channel the appropriate {\it ansatz} reads
\begin{eqnarray}
P^{\dag}&=&\frac{1}{\rfp}\Phi(r){\hat{\mbox{\boldmath $r$}}}\cdot
\mbox{\boldmath $\tau$}\rho \freq \ ,  \qquad
Q^{\dag}_0=\frac{1}{\rfp}\Psi_0(r)\rho \freq \ , 
\label{pansatz}\\
Q^{\dag}_i&=&\frac{1}{\rfp}\left[i\Psi_1(r){\hat r}_i
+\frac{1}{2}\Psi_2(r)\epsilon_{ijk}{\hat r}_j\tau_k\right]
\rho \freq \ .
\label{pansatz2}
\end{eqnarray}
Note that here $\rho$ refers to a properly normalized spinor 
which describes the isospin of the heavy meson multiplet.
Similarly the {\it ansatz} for the S--wave is given by
\begin{eqnarray}
P^{\dag}&=&\frac{1}{\rfp}\Phi(r)\rho \freq \ , \qquad
Q^{\dag}_0=\frac{1}{\rfp}\Psi_0(r)
{\hat{\mbox{\boldmath $r$}}}\cdot
\mbox{\boldmath $\tau$}\rho \freq \ , 
\label{sansatz} \\
Q^{\dag}_i&=&\frac{1}{\rfp}\left[\Psi_1(r){\hat r}_i
{\hat{\mbox{\boldmath $r$}}}\cdot\mbox{\boldmath $\tau$}
+\Psi_2(r)r{\mbox{\boldmath $\tau$}}\cdot\partial_i
{\hat{\mbox{\boldmath $r$}}}\right]
\rho \freq \ .
\label{sansatz2}
\end{eqnarray}
It should be remarked that the isospin matrices, which multiply
the isospinor $\rho$, have (since there are no unmatched indices)
vanishing grand spin $\mbox{\boldmath $G$}$, which is the vector sum 
of total spin and isospin. The above {\it ans\"atze} are substituted 
in the relativistic Lagrangian (\ref{lagheavy}) and the resulting 
action functionals are listed in appendix A. The variation of these 
functionals yields the associated equations of motion. They are 
numerically integrated by adjusting the energy eigenvalue $\epsilon$ 
so that continuous normalizable configurations are 
obtained\footnote{The normalizability condition is quite restrictive. 
For example, it was shown in ref~\cite{Sch96} to prohibit a 
``pentaquark'' solution which would be extracted from eq (\ref{laghl}) 
in the heavy quark limit.}. This value of $\epsilon$ directly yields 
the binding energy of the heavy mesons. The fact that we have
$U(r=0)=-1$ at the spatial origin causes the angular barrier for the 
P--wave heavy meson to vanish while the S--wave acquires a finite 
one. As a result the P--wave heavy meson is more strongly bound. 

Finally we would like to mention the connection to the heavy quark 
limit. In that case the multiplet $H$ is characterized by a single 
radial function \cite{Sch95}. This implies that in the limit 
$M=M^*\to\infty$ the radial functions, which parametrize the bound 
heavy mesons, have to satisfy the linear relations
\begin{eqnarray}
\Psi_1&=&-\Phi, \quad \Psi_2=-2\Phi
\hspace{1cm} {\rm (P-wave)} \ ,
\label{pwhl} \\
\Psi_1&=&-\Phi, \quad \Psi_2=\Phi
\hspace{1.56cm} {\rm (S-wave)}\ ,
\label{swhl}
\end{eqnarray}
together with $\Psi_0=0$ in both cases. Indeed the numerical 
solutions confirm these relations as the heavy meson masses approach 
infinity \cite{Sch96}. It should be remarked that commonly more 
than one bound state exists in each channel. Here we will concentrate
mainly on the lowest one, which is characterized by the radial 
functions having no nodes away from the boundaries $r=0$ and 
$r\to\infty$. However, in special instances we will also discuss
the first radially excited state.

\bigskip
\leftline{\large \it 2.4. Normalization}
\smallskip

As the relativistic Lagrangian (\ref{lagheavy}) is bilinear 
in the heavy meson fields the resulting equations of motion 
are linear. Hence the overall magnitude of the solution 
is not fixed by the equation of motion. Nevertheless, the
equations of motion for the heavy meson fields allow us 
to extract a metric for a scalar product between different 
bound states. In particular its diagonal elements serve 
to properly normalize the bound state wave--functions.
The Lagrange function which results from substituting the 
{\it ans\"atze} (\ref{lightan}) and 
(\ref{pansatz})--(\ref{sansatz}) may generally be written as 
\begin{eqnarray}
L=- M_{\rm cl}\left[F,G,\omega\right]+
I_\epsilon\left[F,G,\omega;\Phi,\Psi_0,\Psi_1,\Psi_2\right]
\rho^{\dag}(\epsilon)\rho(\epsilon) \ .
\label{laggen}
\end{eqnarray}
Here $M_{\rm cl}$ denotes the soliton mass \cite{Ja88} whose 
minimum determines the light meson profiles $F,G$ and $\omega$.
The explicit expressions for the functionals $I_\epsilon$ 
are given in appendix A. The subscript refers to the explicit 
dependence on the energy eigenvalues. Upon canonical quantization 
the Fourier amplitudes $\rho(\epsilon)$ and $\rho^{\dag}(\epsilon)$ 
are respectively elevated to annihilation and creation operators for 
a heavy meson bound state with the energy eigenvalue $\epsilon$. 
Demanding that each occupation of the bound state adds the amount 
$|\epsilon|$ to the total energy yields the normalization 
condition\footnote{In the case that the explicit dependence on 
$\epsilon$ is quadratic the proof is sketched in section 3 of 
ref~\cite{We95}. A simple verification of eq~(\ref{normg}) may be 
also obtained from its close connection to the Noether charge for the 
heavy quark number conservation. The latter is gotten by transforming 
the heavy fields as $P\rightarrow e^{-i\eta(x)}P$,
$Q_\mu\rightarrow e^{-i\eta(x)}Q_\mu$ and computing the quantity
$N= \int d^3x \left.
{\delta{\cal L}(e^{-i\eta}P,\ldots)}/{\delta\partial_0\eta}\ 
\right\vert_{\eta=\partial_0\eta=0}$.
However because the present {\it ansatz} is of the form
$P=\tilde{P}(\mbox{\boldmath$x$})e^{-i\epsilon t}$ 
we may equivalently compute this as
$N= \int d^3x \left.
{\delta{\cal L}\left(e^{-i(\eta+\epsilon t)}
  \tilde{P}\left(\mbox{\boldmath$x$}\right),\ldots\right)}/
{\delta\epsilon}\ \right\vert_{\eta=\partial_0\eta=0}$.
In the one heavy quark subspace this yields 
$1 = {\partial I_\epsilon}/{\partial \epsilon}$. }
\begin{eqnarray}
\Big|\frac{\partial}{\partial\epsilon}
I_\epsilon\left[\Phi,\Psi_0,\Psi_1,\Psi_2\right]\Big|=1
\label{normg}
\end{eqnarray}
in addition to the canonical commutation relation
$[\rho_i(\epsilon),\rho^{\dag}_j(\epsilon^\prime)]=
\delta_{ij}\delta_{\epsilon,\epsilon^\prime}$. Note that
for bound states the energy eigenvalues are discretized. For 
the P--wave channel we obtain the normalization condition
\begin{eqnarray}
&&\Bigg| \int dr r^2 \Bigg\{2\eal\Phi^2
-2\left[\Psi_0^\prime-\eal\Psi_1\right]\Psi_1
+\left[\frac{1}{r}R_\alpha\Psi_0+\eal\Psi_2\right]\Psi_2
\nonumber \\* && \hspace{1cm}
-d\left[\frac{2}{r}{\rm sin}F\Psi_1
-\frac{1}{2}F^\prime\Psi_2\right]\Psi_2
+\frac{4\sqrt{2}c}{gm_V}\frac{1}{r^2}
\left[G\left(G+2\right)\Psi_1-G^\prime r\Psi_2\right]\Phi
\Bigg\} \Bigg|=1
\label{normp}
\end{eqnarray}
from eq (\ref{normg}). For convenience we have employed the abbreviation
$R_\alpha={\rm cos}F-1+\alpha\left(1+G-{\rm cos}F\right)$. Similarly 
for the S--wave channel the condition (\ref{normg}) yields
\begin{eqnarray}
&&\Bigg| \int dr r^2 \Bigg\{2\eal\Phi^2
-2\left[\Psi_0^\prime-\eal\Psi_1\right]\Psi_1
\nonumber \\* && \hspace{1cm}
+4\left[\eal\Psi_2-\frac{R_\alpha+2}{2r}\Psi_0\right]\Psi_2
-2d\left[\frac{2}{r}{\rm sin}F\Psi_1
+F^\prime\Psi_2\right]\Psi_2
\nonumber \\* && \hspace{1cm}
+\frac{4\sqrt{2}c}{gm_V}\frac{1}{r^2}
\left[G\left(G+2\right)\Psi_1+2G^\prime r\Psi_2\right]\Phi
\Bigg\} \Bigg|=1\ .
\label{norms}
\end{eqnarray}

The radial profiles associated with these normalizations
are displayed in figure \ref{fig_1} for the choice $\alpha=-0.3$.
The parameters in the relativistic Lagrangian (\ref{lagheavy}) have 
been set to the charm sector (\ref{heavypara}). The parameters 
entering the light meson Lagrangian (\ref{lagnorm},\ref{lagannorm}) 
are given in eq (\ref{lightpara}).
\begin{figure}
\centerline{
\epsfig{figure=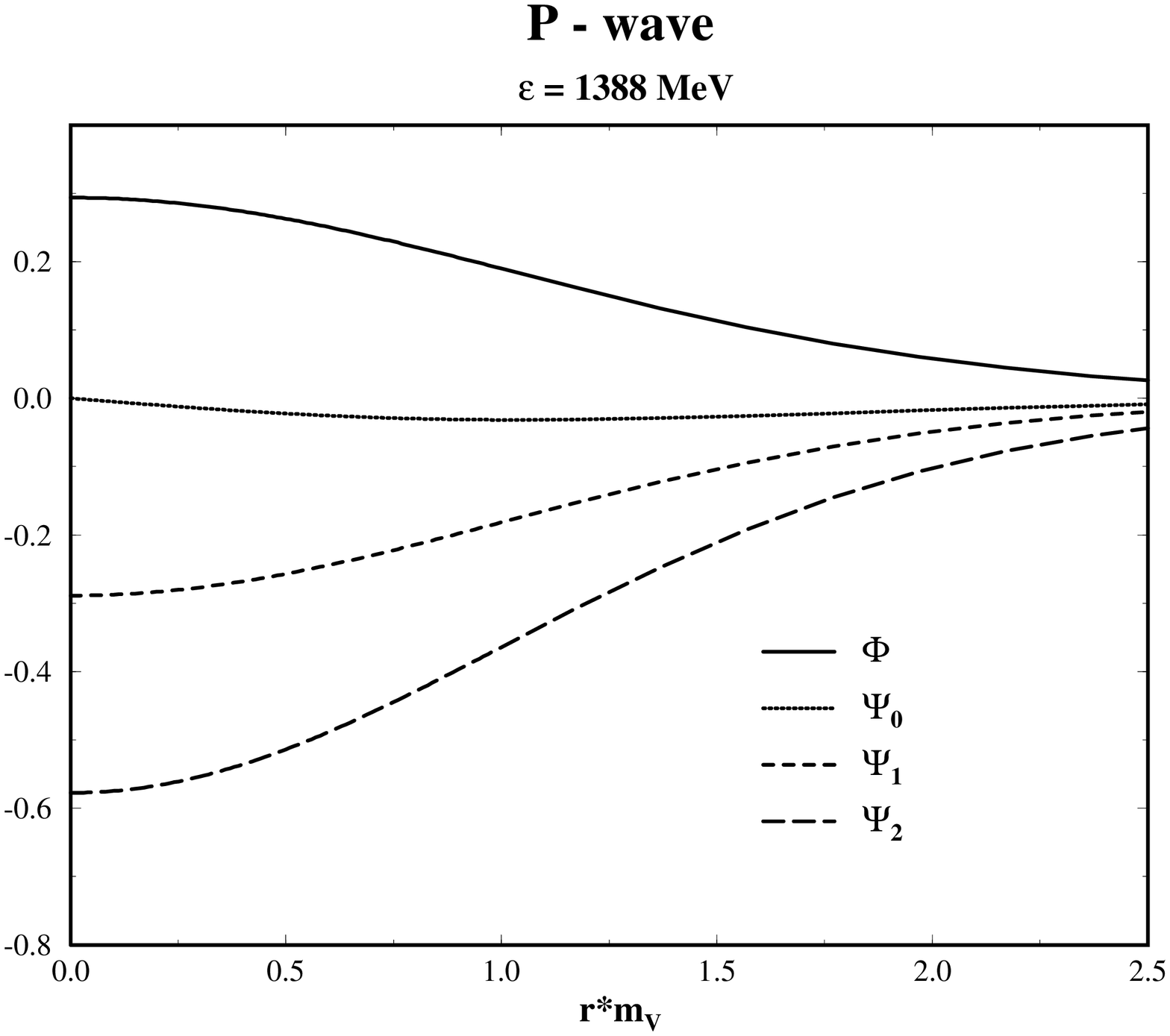,height=7.5cm,width=7.0cm,angle=-90}
\hspace{0.9cm}
\epsfig{figure=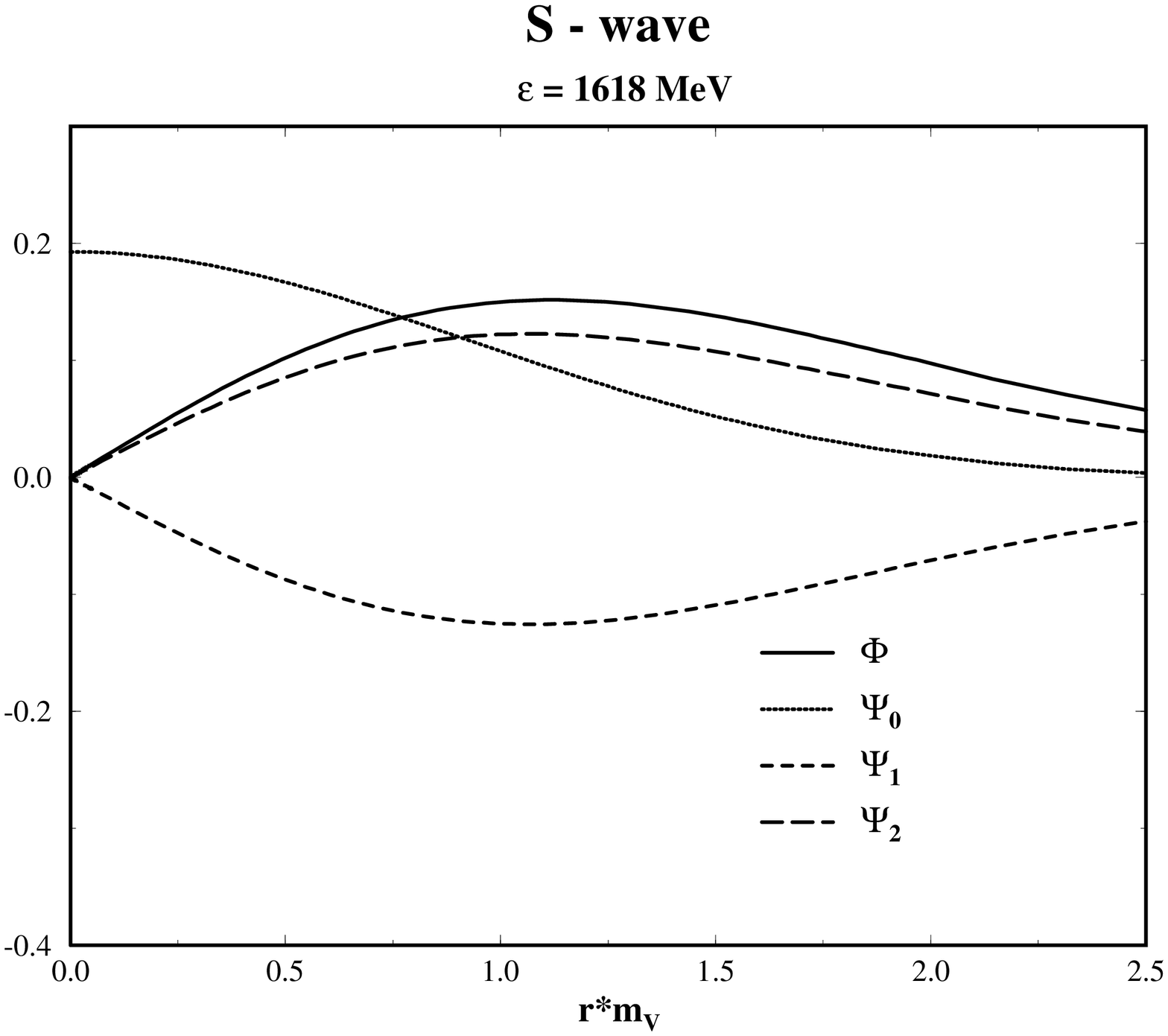,height=7.5cm,width=7.0cm,angle=-90}}
~
\vskip0.1cm
\caption{\label{fig_1}\baselineskip16pt
The profile functions for the bound state wave--functions in the 
P--wave (left panel) and S--wave (right panel) channels. These 
functions are measured in units of $m_V=773{\rm MeV}$. See text for 
the specification of the remaining parameters.}
\end{figure}

\bigskip
\stepcounter{chapter}
\leftline{\Large \it 3. Cranking the Bound Heavy Meson State}
\smallskip

It can easily be verified that the field configurations for both 
the light mesons (\ref{lightan}) and the heavy mesons 
(\ref{pansatz})--(\ref{sansatz2}) are neither eigenfunctions of the 
spin-- nor the isospin generators. Hence these configurations do not 
possess the correct quantum numbers. In order to generate states 
which correspond to physical baryons a cranking procedure is employed. 
In the first step collective coordinates, which parametrize the 
(iso--) spin orientation of the meson configuration, are introduced.

\bigskip
\leftline{\large \it 3.1. Collective Coordinates}
\smallskip

Time--dependent solutions to the equations of motion are 
required to obtain non--vanish\-ing spin and isospin as the 
corresponding Noether charges. Unfortunately these solutions are 
unknown. Taking, however, into account that static rotations in 
coordinate and isospin spaces do not change the potential part of 
the Lagrange function, the assumption that these rotations are 
time--dependent seems to be a reasonable approximation. We therefore 
extend the soliton {\it ansatz} (\ref{lightan}) by
\begin{eqnarray}
\xi\longrightarrow A(t)\xi A^{\dag}(t)
\quad {\rm and} \quad 
\rho_\mu\longrightarrow A(t) \rho_\mu A^{\dag}(t) \ .
\label{colrot}
\end{eqnarray}
Note that $\rho_\mu$ contains both isoscalar and isovector pieces.
It should also be remarked that introducing only an isospin 
rotation as in eq (\ref{colrot}) is sufficient because the 
hedgehog structure of the classical configuration (\ref{lightan})
allows one to express a spatial rotation as one in isospin 
space. The time--dependence of the collective rotations is 
most conveniently parametrized by introducing angular 
velocities $\mbox{\boldmath $\Omega$}$ via
\begin{eqnarray}
A^{\dag}(t)\frac{d}{dt} A(t)=\frac{i}{2}
\mbox{\boldmath $\tau$}\cdot
\mbox{\boldmath $\Omega$} \ .
\end{eqnarray}
In the specific case that the Lagrangian contains terms which are 
linear in the time derivative as in eq (\ref{lagannorm}), additional 
field components are induced. For the light vector mesons these 
are linear in the angular velocities
\begin{eqnarray}
\omega_i=\frac{2}{r}\varphi(r)\epsilon_{ijk}\Omega_j{\hat r}_k
\quad {\rm and}\quad
\rho_0^k=\xi_1(r)\Omega_k+\xi_2(r){\hat{\mbox{\boldmath $r$}}}\cdot
\mbox{\boldmath $\Omega$}{\hat r}_k \ .
\label{induced}
\end{eqnarray}
Substituting the configurations (\ref{colrot}) and (\ref{induced})
into the light meson Lagrangian yields a term which is quadratic in 
the angular velocities. The constant of proportionality defines 
the moment of inertia 
$\alpha^2\left[F,G,\omega;\xi_1,\xi_2,\varphi\right]$. The 
induced radial functions $\varphi(r)$, $\xi_1(r)$ and $\xi_2(r)$ are 
obtained from a variational approach to $\alpha^2$ 
\cite{Me89}\footnote{The term $\alpha^2$ refers to a frequently
adopted notation for the moment of inertia and should not be 
confused with the coupling constant $\alpha$ in eq (\ref{covderp}).}. 
The resulting equations of motion are coupled inhomogeneous 
differential equations with the classical fields $F,G$ and $\omega$, 
which are fixed from extremizing the classical mass, acting as 
sources. Here it is worth mentioning that $\alpha^2$ is of the 
order $N_C$.

Since the heavy mesons are also isospinors the collective rotation 
has to be applied as well. In analogy to eq (\ref{colrot}) we write
\begin{eqnarray}
P^{\dag}\longrightarrow A(t)P^{\dag}
\quad {\rm and} \quad
Q^{\dag}_\mu\longrightarrow A(t)Q^{\dag}_\mu \ .
\label{rotheavy}
\end{eqnarray}
Substituting the collectively rotating configurations 
into the total Lagrangian finally yields
\begin{eqnarray}
L_P=-M_{\rm cl}+I_\epsilon^{(P)}\rho^{\dag}\rho
+\frac{1}{2}\alpha^2 {\mbox{\boldmath $\Omega$}}^2
+\frac{1}{2}\chi_P \rho^{\dag}\mbox{\boldmath $\Omega$}\cdot
\mbox{\boldmath $\tau$}\rho \ .
\label{collp}
\end{eqnarray}
For convenience we have omitted  the argument of the iso--spinor 
$\rho$. The classical mass $M_{\rm cl}$, which upon minimization 
provides the soliton profiles (\ref{lightan}), and the moment of 
inertia $\alpha^2$
are functionals of only the light meson fields. The quantity 
$I_\epsilon^{(P)}$ is given in eq (\ref{pwlag})
and has already been employed to obtain the bound state P--wave 
profiles (\ref{pansatz}). The new quantity is the hyperfine 
parameter $\chi_P$ whose explicit expression is displayed in 
appendix B (\ref{eq3}). 

\bigskip
\leftline{\large \it 3.2. Quantization of the 
Collective Coordinates}

\smallskip

Here we will discuss how the canonical quantization of the 
collective coordinates $A$ leads to states which may be 
identified with physical baryons. In order to construct 
Noether charges we first have to consider the variation 
of the fields under infinitesimal symmetry transformations.
For the isospin transformation we observe
\begin{eqnarray}
\left[\phi,i\frac{\tau_i}{2}\right]
=-D_{ij}(A)\frac{\partial{\dot \phi}}{\partial \Omega_j} 
+ \ldots \ .
\label{isorot}
\end{eqnarray}
Here $\phi$ refers to any of the iso--rotating meson fields and 
the ellipses represent terms, which are subleading in $1/N_C$, as 
{\it e.g.} time derivatives of the angular velocities which
might arise from eq~(\ref{induced}).
Furthermore
$D_{ij}(A)=(1/2)\ {\rm tr}(\tau_i A \tau_j A^{\dag})$ denotes the 
adjoint representation of the collective rotations $A$. From 
eq (\ref{isorot}) we conclude that the total isospin is related 
to the derivative of the Lagrange function with respect to the 
angular velocities
\begin{eqnarray}
I_i=-D_{ij}(A)\frac{\partial L_P}{\partial \Omega_j} \ .
\label{isospin}
\end{eqnarray} 
Next we note that the total spin operator $\mbox{\boldmath $J$}$
enters the grand spin operator $\mbox{\boldmath $G$}$ in the 
laboratory frame via
\begin{eqnarray}
G_i=J_i+D_{ij}^{-1}(A)I_j = J_i - J^{\rm sol}_i\ .
\label{gspin1}
\end{eqnarray}
For convenience we have defined the spin carried by the soliton 
$\mbox{\boldmath $J$}^{\rm sol}=
\partial L_P / \partial \mbox{\boldmath $\Omega$}$. As a consequence 
of the relation (\ref{isospin}), its absolute value is identical to 
that of the isospin, {\it i.e.}
$(\mbox{\boldmath $J$}^{\rm sol})^2=\mbox{\boldmath $I$}^2=I(I+1)$.
By construction the light meson fields do not contribute 
to $\mbox{\boldmath $G$}$. Even more importantly and as has been 
noted before, the pieces of the heavy meson wave--functions 
(\ref{rotheavy}), which multiply the spinor $A\rho$, have zero grand 
spin too. Using the normalization condition (\ref{normp}) one 
therefore ends up with
\begin{eqnarray}
\mbox{\boldmath $G$}=-\rho^{\dag}
\frac{\mbox{\boldmath $\tau$}}{2}\rho \ .
\label{gspin2}
\end{eqnarray}
This relation will be helpful because it relates the operator 
multiplying the hyperfine parameter in the collective Lagrangian
(\ref{collp}) to the spin and isospin operators. The collective 
piece of the Hamiltonian is obtained from the Legendre transformation
\begin{eqnarray}
H_P^{\rm coll}=\mbox{\boldmath $\Omega$}
\cdot\mbox{\boldmath $J$}^{\rm sol}-L_P^{\rm coll}
=\frac{1}{2\alpha^2}\left[\mbox{\boldmath $J$}^{\rm sol} 
+\chi_P \mbox{\boldmath $G$}\right]^2\ ,
\label{collham}
\end{eqnarray}
Here $L_P^{\rm coll}$ refers to the $\mbox{\boldmath $\Omega$}$
dependent terms in eq (\ref{collp}). Finally the mass formula for 
an even parity baryon with a single heavy quark becomes
\begin{eqnarray}
M_P=M_{\rm cl}+|\epsilon_P|
+\frac{1}{2\alpha^2}\left[\chi_P J(J+1)+(1-\chi_P)I(I+1)\right] \ ,
\label{massp}
\end{eqnarray}
where contributions of ${\cal O}(\chi_P^2)$, which apparently  
are quartic in the heavy meson wave--function, have been omitted 
for consistency because terms of that order have been excluded from 
the very beginning. Also the omitted terms are subleading in the 
$1/M$ expansion since $\chi_P$ goes as $1/M$, {\it cf.}
figure \ref{fig_2} and Appendix B. In addition, the operator
contained in the omitted term,
$\left(\rho^{\dag}\mbox{\boldmath $\tau$}\rho\right)\cdot
\left(\rho^{\dag}\mbox{\boldmath $\tau$}\rho\right)$, does
not contribute to the hyperfine splitting. The reason is that
from canonical commutation relations for the
components of the isospinor $\rho$,
$\left[\rho_i,\rho_j^{\dag}\right]=\delta_{ij}$, this operator
is shown to be $N_Q\left(N_Q+2\right)$, where $N_Q$ is the
occupation number for the heavy meson bound state \cite{Ca85}.
Hence this term contains neither the spin nor the isospin
quantum numbers.

{}From eq (\ref{massp}) we recognize that the spin degeneracy 
between baryons containing a heavy quark vanishes in the heavy quark 
limit because $\chi_P$ approaches zero. Of course, this result is a 
direct consequence of the spin--flavor symmetry and would not have 
come out in case the various Lorentz and chirally invariant terms in 
eq (\ref{lagheavy}) had been chosen arbitrarily.

\bigskip
\leftline{\Large \it 3.3 The Odd Parity State}
\smallskip

The S-- and P--wave heavy channels decouple because 
they have opposite parity. Therefore the quantization of the 
S--wave bound state may be considered independently from
the P--wave case, which has been discussed in the 
preceding section. The calculation, which proceeds 
along the lines of the one discussed in subsection 3.1, yields
\be
L=-M_{\rm cl}+I_\epsilon^{(S)}\rho^{\dag}\rho
+\frac{1}{2}\alpha^2 {\mbox{\boldmath $\Omega$}}^2
+\frac{1}{2}\chi_S \rho^{\dag}\mbox{\boldmath $\Omega$}\cdot
\mbox{\boldmath $\tau$}\rho \ .
\label{colls}
\ee
Here, of course, the spinor $\rho$ corresponds to the one of the 
bound heavy meson in the S-wave channel. The explicit expression 
for the corresponding hyperfine parameter $\chi_S$ is given in 
eq~(\ref{eq5}) in Appendix B.

We may apply the same quantization procedure as for the P--wave. 
This results in the mass formula 
\begin{eqnarray}
M_S=M_{\rm cl}+|\epsilon_S|
+\frac{1}{2\alpha^2}\left[\chi_S J(J+1)+(1-\chi_S)I(I+1)\right] 
\label{masss}
\end{eqnarray}
for baryons constructed as a light baryon and a single occupation
of the bound state for the heavy meson being in the S--wave channel. 

Let us add a brief comment on the $1/N_C$ dependences in eqs
(\ref{massp}) and (\ref{masss}). The classical mass is 
${\cal O}(N_C)$ while in leading order the binding energies
are ${\cal O}(N_C^0)$. As already noted above, the moment of 
inertia is ${\cal O}(N_C)$ while $\chi\sim{\cal O}(N_C^0)$. 
Therefore the hyperfine splitting is not only subleading in 
the heavy quark limit but also in the $1/N_C$ expansion.

\bigskip
\stepcounter{chapter}
\leftline{\large \it 4. Numerical Results}
\smallskip

In this section we will discuss the numerical results 
obtained for the masses of the heavy baryons in the model discussed 
above. In particular we will concentrate on the spin and isospin 
splitting in the realistic case of finite heavy meson masses 
(\ref{heavypara}). It should be noted that sizable quantum 
corrections occur for the classical soliton mass $M_{\rm cl}$ 
\cite{Me96}. It seems that these corrections are (approximately) equal 
for all baryons. Hence we will only consider mass differences between 
various baryons. In that case the absolute value of the classical mass 
$M_{\rm cl}$ is redundant. The parameters in the light sector cannot 
completely be determined from properties of the corresponding mesons. 
The remaining (limited) parameter space is, however, more than fully 
constrained by a best fit to the mass differences of the low--lying 
$\frac{1}{2}^+$ and $\frac{3}{2}^+$ baryons in the light sector. 
This yields: 
\begin{eqnarray}
g=5.57,\quad && m_V=773\,{\rm MeV}
\nonumber \\
\gamma_1=0.3, \quad \gamma_2&=&1.8, \quad \gamma_3=1.2 \ .
\label{lightpara}
\end{eqnarray}
The resulting mass differences for the light baryons all agree within
about 10\% \cite{Pa91}. The corresponding moment of inertia is 
$\alpha^2=5.00{\rm GeV}^{-1}$.  In eqs (\ref{massp}) and (\ref{masss})
$1/\alpha^2$ enters as the coefficient of those terms which
determine the hyperfine splitting. Hence a fine--tuning of the
parameters (\ref{lightpara}) to {\it e.g.}
$\alpha^2=5.11{\rm GeV}^{-1}$, which exactly reproduces the
$\Delta$--nucleon mass difference, has only negligible effects
on the predicted hyperfine splittings.

Before discussing the implications for the physical parameter results 
(\ref{heavypara}) we would like to comment on the heavy limit
behavior of the hyperfine splitting parameters $\chi_P$ and $\chi_S$.
For this purpose we have plotted these quantities as functions of 
$M=M^*$ in figure \ref{fig_2}. We see that for both channels the 
splitting parameters decrease when the heavy limit is approached. In 
the appendix we show that the leading order term in the heavy quark 
expansion indeed is proportional to $1/M$. 
\begin{figure}
\centerline{
\epsfig{figure=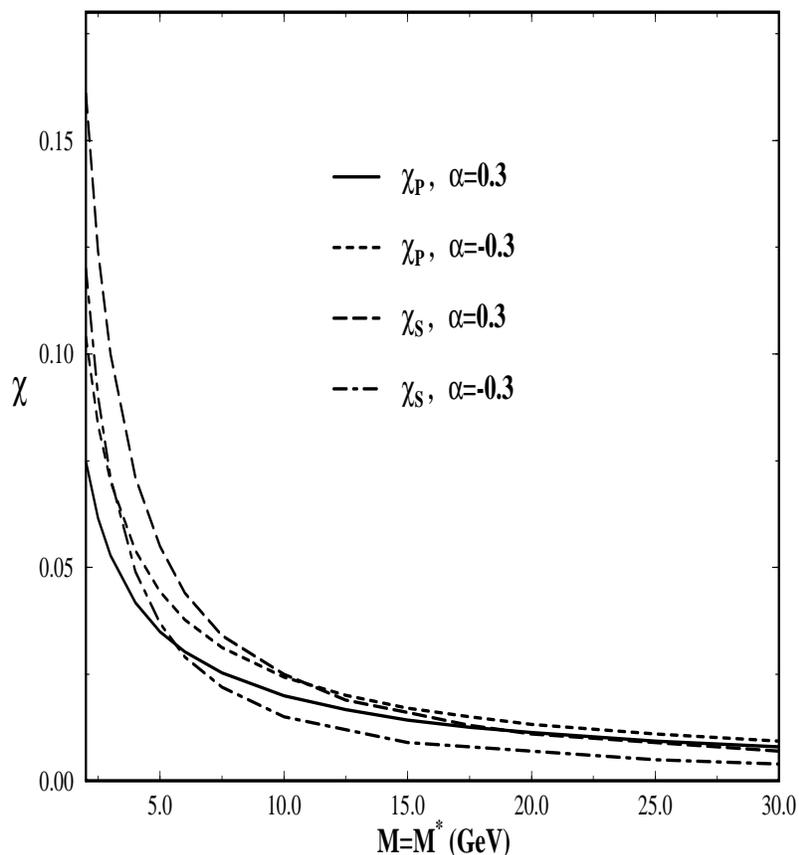,height=9.0cm,width=12.5cm,angle=-90}}
~\vskip0.3cm
\caption{\label{fig_2}\baselineskip16pt
The hyperfine splitting parameters $\chi_P$ and $\chi_S$ 
a functions of identical heavy meson masses for different values 
of the coupling constant $\alpha$.}
\end{figure}
Clearly the hyperfine parameter in the S--wave channel decreases 
somewhat more quickly with the heavy meson mass than in the 
P--wave channel.

For fixed $M$ the hyperfine parameters in the two channels behave 
oppositely with regard to the undetermined coupling constant $\alpha$: 
While $\chi_P$ decreases when $\alpha$ becomes larger, $\chi_S$ 
increases. In ref \cite{Ha97} we have shown that a major fraction of 
the P--wave hyperfine constant is due to terms in the relativistic 
Lagrangian (\ref{lagheavy}) which do not manifestly break the heavy 
spin symmetry rather than to terms, which explicitly break this 
symmetry; as for example $M\ne M^*$. For a quantitative discussion of 
this hidden contribution we perform the calculation using identical 
masses from the charm sector {\it i.e.} $M=M^*=1.865{\rm GeV}$ and 
furthermore $\alpha=0.3$. We take all other parameters as in eq 
(\ref{heavypara}). This results in $\chi_P=0.080$. From table 
\ref{tab_1} we recognize that this is about 80\% of the value 
obtained using the physical masses $M=1.865{\rm GeV}$ and 
$M^*=2.007{\rm GeV}$. In the case of the S--wave the hidden piece is 
even more dominant. For the symmetric choice of the mass parameters 
one finds $\chi_S=0.175$ which is more than 90\% of the value 
displayed in table \ref{tab_1} for $\alpha=0.3$. It is also 
interesting to compare figure 4.1 with the corresponding curve in 
figure 1 of ref~\cite{Ha97}, pertaining to the model without light 
vectors. This makes it clear that the light vector model predicts a 
substantially larger $\chi_P$.

In section 5, some more discussion of the ``hidden'' hyperfine 
splitting terms is given.  In addition, three more ``manifest'' 
heavy spin symmetry violating terms associated with the relativistic 
Lagrangian (\ref{lagheavy}) are treated in the perturbative expansion.
Since the manifest $(M^{\ast}-M)$ contribution is relatively small
it is reasonable to expect that the others will be small too.

Let us next discuss the spectrum of the baryons containing a single 
heavy quark. For this case we assume the realistic masses as in 
eq (\ref{heavypara}). In table \ref{tab_1} the numerical results for 
the lowest S-- and P--wave bound states in the charm sector are 
displayed. As already noted in ref \cite{Sch96} the binding energy 
\begin{eqnarray}
\omega = M - |\epsilon|
\label{bindeng}
\end{eqnarray}
decreases with growing coupling constant $\alpha$. This is the case 
for both the P-- and S--wave channels. For $M\rightarrow\infty$ the 
heavy limit \cite{Sch95}
\begin{eqnarray}
\omega\longrightarrow\frac{3}{2}dF^\prime(0)
+\frac{3\sqrt{2}c}{gm_V}G^{\prime\prime}(0)
-\frac{\alpha}{2}\omega(0)
\label{ebpw}
\end{eqnarray}
will be attained\footnote{Note that the conventions in 
ref~\cite{Sch95} differ from the present ones as explained in 
Appendix B of ref~\cite{Sch96}.}. As in the discussion of figure 
\ref{fig_2} we see that the hyperfine parameters in these two channels 
behave oppositely as functions of $\alpha$.
\begin{table}
\caption{\label{tab_1}\baselineskip18pt
Parameters for heavy baryons and mass differences 
with respect to $\Lambda_c$. Primes indicate negative parity
baryons, {\it i.e.} S--wave bound states. All energies are in MeV.}
~
\newline
\centerline{
\begin{tabular}{l | c c c c c | c | c}
$\alpha$ & -0.1 & 0.0 & 0.1 & 0.2 & 0.3 & Expt. & Skyrme \\
\hline
$\omega_P$    & 564 & 544 & 522 & 500 & 478 & & 243\\
$\chi_P$      & 0.147 & 0.140 & 0.131 & 0.123 & 0.114 & & 0.053 \\
$\omega_S$    & 316 & 298 & 281 & 264 & 247 & & 57\\
$\chi_S$      & 0.172 & 0.181 & 0.189 & 0.197 & 0.205 & &0.346 \\  
\hline
$\Sigma_c$     & 171 & 172 & 174 & 175 & 177 & 168 & 185\\
$\Sigma_c^*$   & 215 & 214 & 213 & 212 & 211 & 233 & 201\\
$\Lambda_c^\prime $ & 250 & 249 & 245 & 242 & 238 & 308 & 208 \\
$\Sigma_c^\prime $ & 415 & 413 & 408 & 402 & 397 & ? & 335\\
$\Sigma_c^{\prime *} $ & 468 & 467 & 464 & 461 & 458 & ? & 437 \\
\hline
$N$ & -1237 & -1257 & -1278 & -1299 & -1321 & -1345 & -1553\\
\hline
$\Lambda_b$ & 3160 & 3164 & 3167 & 3170 & 3173 & $3356\pm50$
&3215 
\end{tabular}
}
\end{table}
Here we have chosen to measure the mass differences with respect to 
the lightest charmed baryon, $\Lambda_c$. Hence the mass differences
with respect to $\Sigma_c$ and $\Sigma_c^*$ directly reflect the 
$\alpha$--dependence of hyperfine parameter $\chi_P$ while the 
corresponding dependence of the binding energy $\omega_P$ can be 
extracted from the splitting relative to the nucleon. In addition 
the splitting with respect to the negative parity charmed baryons 
reflects the $\alpha$--dependence of the S--wave channel 
binding energy $\omega_S$. Finally the mass difference to 
$\Lambda_b$ contains the energy eigenvalues and hyperfine 
parameters computed with the $B$ and $B^*$ meson masses in eq
(\ref{heavypara}).

While the mass difference to the nucleon is improved with
a positive value for $\alpha$, the agreement for the 
mass differences between the heavy baryons slightly 
deteriorates when increasing this parameter. Nevertheless,
fair agreement with the experimental data is 
achieved for quite a range of $\alpha$.

Table \ref{tab_1} also contains the model predictions when 
the background soliton is taken from the basic Skyrme model 
\cite{Sk61,Ad83} which does not include the light vector mesons. 
Here we have adjusted the only free parameter 
($e_{\rm Skyrme}=4.25$) to reproduce the $\Delta$--nucleon mass 
difference. From the $\Lambda_c$--nucleon mass difference we observe 
that in comparison with the nucleon the masses of the heavy baryons 
are predicted about $200{\rm MeV}$ too large. This confirms the 
above statement that the spectra of both the light and the heavy 
baryons can only be reasonably reproduced when light vector mesons 
are included. This conclusion can already be drawn from the too 
small binding energies \cite{Sch96}. The hyperfine corrections 
make only minor changes in the $\Lambda_b$--$\Lambda_c$ splitting.

In table \ref{tab_2} we display the analogous predictions for 
the bottom sector. According to the heavy spin symmetry the 
binding energies of the P-- and S--wave channels approach each 
other. Hence the mass differences between the even and odd parity 
baryons containing a bottom quark correspondingly decrease. As was 
already inferred from figure \ref{fig_2} we confirm upon comparison 
with table \ref{tab_1} that $\chi_P$ decreases less quickly 
with the heavy meson mass than $\chi_S$.
\begin{table}
\caption{\label{tab_2}\baselineskip18pt
Parameters for heavy baryons and mass differences 
with respect to $\Lambda_b$. Primes indicate negative parity
baryons, {\it i.e.} S--wave bound states. All energies are in MeV.
The empirical value for the relative position of the nucleon
is $4701\pm50{\rm MeV}$ [31].}
~
\newline
\centerline{
\begin{tabular}{l | c c c c c }
$\alpha$ & -0.1 & 0.0 & 0.1 & 0.2 & 0.3 \\
\hline
$\omega_P$    & 811 & 786 & 762 & 737 & 713 \\
$\chi_P$      & 0.055 & 0.053 & 0.050 & 0.048 & 0.045 \\
$\omega_S$    & 639 & 617 & 595 & 573 & 552 \\
$\chi_S$      & 0.043 & 0.046 & 0.049 & 0.052 & 0.055 \\  
\hline
$\Sigma_b$     & 189 & 189 & 190 & 190 & 191  \\
$\Sigma_b^*$   & 206 & 205 & 205 & 205 & 205  \\
$\Lambda_b^\prime $ & 171 & 168 & 167 & 164 & 161  \\
$\Sigma_b^\prime $ & 363 & 359 & 358 & 354 & 351  \\
$\Sigma_b^{\prime *} $ & 375 & 373 & 371 & 369 & 367  \\
\hline
$N$ & -4397 & -4422 & -4446 & -4471 & -4494
\end{tabular}
}
\end{table}
Except for $\Lambda_b$ no empirical data for the masses of these 
baryons are known at present. These results for the mass differences 
among the bottom baryons are predictions of the model which can, in 
the future, be compared with experiment. As could have been inferred 
from the next to last row in table \ref{tab_1} the absolute position 
of the bottom multiplet is about $200\pm50{\rm MeV}$ too low. 
On the absolute scale this apparently is only a 5\% deviation from 
the data. Certainly a larger value $\alpha\approx1$,
which corresponds to a model for light vector resonance dominance 
of the heavy meson form factor~\cite{Ja95}, would yield an 
excellent agreement for the mass difference between $\Lambda_b$ 
and the nucleon. On the other hand such a choice would slightly
spoil the nice picture for the charm multiplet.

In table \ref{tab_3} we list the numerical results for baryons
constructed from the first radially excited P--wave bound 
state in the charm sector.
\begin{table}
\caption{\label{tab_3}\baselineskip18pt
Parameters for radially excited heavy baryons and mass differences
with respect to $\Lambda_c$. A tilde refers to a radially 
excited P--wave baryon. All energies are in MeV.}
~
\newline
\centerline{
\begin{tabular}{l | c c c c c }
$\alpha$ & -0.1 & 0.0 & 0.1 & 0.2 & 0.3 \\
\hline
$\omega_P$    & 125 & 113 & 101 &  90 &  79  \\
$\chi_P$      & 0.020 & 0.014 & 0.007 & 0.000 &-0.007  \\
\hline
$\tilde{\Lambda}_c$ & 429 & 419 & 409 & 398 & 388   \\
$\tilde{\Sigma}_c$  & 625 & 618 & 607 & 598 & 589   
\end{tabular}
}
\end{table}
The particle data group (PDG) lists an excited $\Lambda_c(2625)$,
about $340{\rm MeV}$ above the $\Lambda_c$ \cite{PDG96}, 
although this state is more likely, according to the PDG, to have 
$J^P=\frac{3}{2}^-$. In the bound state picture this would 
require a D--wave ($G=3/2$) heavy meson bound to the soliton. Here we 
have not discussed that channel. A preliminary discussion in the 
perturbative approach is given in the next section.

The preceding calculations are based on the 
$N_{\rm C}\rightarrow\infty$ limit in which the nucleon is 
infinitely heavy. From a common sense point of view this is 
peculiar since the nucleon is actually lighter than the heavy 
mesons being bound to it in the model. 
Hence, for comparison with experiment it is desirable to estimate
kinematic effects associated with the nucleon's motion.
These are expected~\cite{Sch95} to lower the binding energy of 
the heavy baryons which have up to now come out too high
(see $\Lambda_c$--$N$ mass difference in Table~4.1, for example.).
In order to estimate these kinematical effects in the bound state 
approach we have substituted the reduced masses 
\begin{eqnarray}
\frac{1}{M}\longrightarrow\frac{1}{M_{\rm cl}}
+\frac{1}{M}
\quad {\rm and} \quad
\frac{1}{M^*}\longrightarrow\frac{1}{M_{\rm cl}}
+\frac{1}{M^*}
\label{reduced}
\end{eqnarray}
into the bound state equations. In a non--relativistic treatment
this corresponds to the elimination of the center of mass motion 
\cite{Sch95}. 
The results for the spectrum of the heavy baryons 
obtained from the replacement (\ref{reduced}) are in displayed in 
table \ref{tab_4}. 
\begin{table}
\caption{\label{tab_4}\baselineskip18pt
Parameters for heavy baryons and mass differences with 
respect to $\Lambda_c$. Primes indicate negative parity states, 
{\it i.e.} S--wave bound states. All energies are in MeV. In this 
calculation the reduced masses (4.4) enter the bound state 
equations from which the binding energies are extracted. The 
physical meson masses 1865MeV and 5279MeV are used when computing 
the mass differences to the nucleon and the $\Lambda_b$ from these 
binding energies. Radially excited states are omitted because 
they are only very loosely bound, if at all.
The empirical data are taken from the PDG [31], see also [32].}
~
\newline
\centerline{
\begin{tabular}{l | c c c c c c c c | c}
$\alpha$ & 0.0 & -0.1 & -0.2 & -0.3 & -0.4 & -0.5 & -0.6 & -0.7 & Expt. \\
\hline
$\omega_P$    & 450 & 469 & 488 & 508 & 527 & 546 & 566 & 585 & \\
$\chi_P$      & 0.212 & 0.232 & 0.246 & 0.260 & 0.273 & 0.286 & 
0.299 & 0.312 & \\
$\omega_S$    & 123 & 134 & 146 & 158 & 171 & 184 & 197 & 210 & \\
$\chi_S$      & 0.410 & 0.399 & 0.387 & 0.374 & 0.361 & 0.346 & 
0.331 & 0.315 & \\  
\hline
$\Sigma_c$     & 158 & 154 & 151 & 148 & 145 & 143 & 140 & 138 & 168 \\
$\Sigma_c^*$   & 221 & 223 & 225 & 226 & 227 & 229 & 230 & 231 & 233 \\
$\Lambda_c^\prime $ & 342 & 346 & 353 & 359 & 363 & 367 & 
371 & 375 & 308 \\
$\Sigma_c^\prime $  & 460 & 468 & 475 & 484 & 490 & 497 & 
505 & 512 & ? \\
$\Sigma_c^{\prime *} $ & 583 & 587 & 591 & 596 & 599 & 601 & 
605 & 607 & ? \\
\hline
$N$ & -1356 & -1338 & -1320 & -1302 & -1283 & -1265 & -1246 & 
-1228 & -1345 \\
\hline
$\Lambda_b$ & 3285 & 3282 & 3280 & 3278 & 3275 & 3272 & 3271 & 
3269 & $3356 \pm 50$
\end{tabular}
}
\end{table}
Again we consider $\alpha$ as a free parameter. We notice that 
there is a remarkable improvement in the prediction for the 
$\Lambda_b$ mass, which was previously the worst one. The changes 
in some of the mass parameters can approximately be compensated 
by a suitable re--adjustment of $\alpha$. 
For $\alpha\approx0.0$ to $-0.4$ the agreement with the 
existing data is quite reasonable. When using the reduced meson 
masses the $\Sigma_c$ baryon is always predicted a bit too 
light while it is too heavy when the physical masses are 
substituted in the bound state equation. For $\Lambda_c^\prime$
the situation is opposite. While the use of the physical meson masses
gives too small a mass, the substitution of the reduced masses 
gives too large a prediction for the mass of this baryon. These 
results indicate that kinematical corrections are indeed 
important. It should, however, be remarked that with this
replacement the heavy quark limit $\chi\to0$ cannot be attained
because the mass parameters in the bound state equations will
remain finite and hence the linear relations (\ref{pwhl}) and 
(\ref{swhl}) will not be satisfied. However, these relations 
are essential to verify the limit $\chi\to0$ for $M=M^*\to\infty$.
Nevertheless we think that the replacement (\ref{reduced}) 
provides sensible insight in the relevance of kinematical 
corrections.

It is interesting to see how far the heavy quark approach 
can be pushed to lighter quarks. To answer this question we 
have considered the strange quark. In the corresponding kaon 
sector the P--wave is only very loosely bound when the physical 
masses are substituted. On the other hand sizable binding 
energies are obtained when the reduced masses are used \cite{Sch96}. 
This behavior is somewhat different from the charm and bottom 
sector and can be understood by noting that the difference 
$M^*-M$ is considerably reduced when using (\ref{reduced}). In the 
heavy sectors (charm and bottom) this difference is small in any 
event.
\begin{table}
\caption{\label{tab_5}\baselineskip18pt
Same as table 4.4 for even parity baryons
in the kaon sector.}
~
\newline
\centerline{
\begin{tabular}{l | c c c c c c c c | c}
$\alpha$ & 0.0 & -0.1 & -0.2 & -0.3 & -0.4 & -0.5 & -0.6 & -0.7 & Expt. \\
\hline
$\omega_P$    &  80 &  94 & 109 & 124 & 140 & 155 & 171 & 188 & \\
$\chi_P$      & 0.346 & 0.371 & 0.394 & 0.417 & 0.439 & 0.460 & 
0.479 & 0.498 & \\
\hline
$\Sigma$     & 131 & 126 & 121 & 117 & 112 & 108 & 104 & 100 & 77 \\
$\Sigma^*$   & 235 & 237 & 239 & 242 & 244 & 246 & 248 & 250 & 269 \\
\hline
$N$ & -366 & -354 & -341 & -327 & -313 & -300 & -285 & 
-269 & -177
\end{tabular}
}
\end{table}
The resulting spectrum for the strange baryons is shown in table 
\ref{tab_5}. The comparison with the experimental data shows that 
even the use of the reduced masses does not provide sufficient 
binding. In the S--wave channel the situation is worse, even when 
the reduced masses are substituted bound states are not detected unless
$\alpha\le-1.0$. The failure of the present approach in the
strange sector strongly suggests that for these baryons a 
chirally invariant set--up \cite{We96} is more appropriate.

\bigskip
\stepcounter{chapter}
\leftline{\Large \it 5. The Perturbative Approach}
\smallskip

The perturbative approach can illuminate several aspects of the
hyperfine splitting problem. This is due to the heavy quark symmetry 
which is naturally exploited by making an expansion in powers of 
$1/M$ using the heavy field formalism. Our starting Lagrangian 
(\ref{lagheavy}) has been set up in such a way as to yield a 
heavy quark symmetric result as $M\rightarrow\infty$ when 
$M=M^{\ast}$ is assumed, {\it cf.} eq (\ref{laghl}). The 
perturbative $1/M$ expansion is more general (presumably exact) but 
less predictive.  Thus the $1/M$ expansion provides a useful 
calibration in the large $M$ limit. Since it deals with perturbation 
matrix elements it provides us with a convenient classification of 
the various sources of hyperfine splitting. The method is also 
advantageous in that it can be extended, without too much algebraic 
work, to different channels of interest. On the other hand, once the 
particular channels of interest are settled on, it is clearly more 
convenient to employ the exact numerical solution, which efficiently 
sums up a class of $1/M$ corrections.

The leading order Lagrangian (\ref{laghl}) can be supplemented by 
terms which manifestly break the heavy quark symmetry to leading order
($M^0$ with the present normalization) as follows:
\begin{eqnarray}
\frac{1}{M}{\cal L}'_H &=&
\frac{M-M^{\ast}}{8} \mbox{\rm Tr}
\left[ H \sigma_{\mu\nu} \overline{H} \sigma^{\mu\nu} \right]
+\frac{(d-d')}{2} \mbox{\rm Tr}
\left[ H p_\mu \overline{H} \gamma^\mu \gamma_5 \right]
\nonumber\\
&& \quad
{}-i \frac{\sqrt{2}(c-c')}{m_V} \mbox{\rm Tr}
\left[ \gamma_\mu \gamma_\nu H F^{\mu\nu}(\rho) \overline{H} \right]
+ \tilde{\alpha} V^\beta \mbox{\rm Tr}
\left[
  H \sigma_{\mu\nu} \left( \tilde{g} \rho_\beta - v_\beta \right)
  \overline{H} \sigma^{\mu\nu}
\right]
\ .
\label{heavy next}
\end{eqnarray}
Here the $(M-M^{\ast})$ term measures the heavy spin violation due to
the heavy pseudoscalar -- heavy vector mass difference.
The $(d-d')$ term measures the heavy spin violation induced by choosing
different coefficients for the fifth and sixth terms in
eq (\ref{lagheavy}) (or see eq (2) in ref \cite{Ha97}) while the
$(c-c')$ term  corresponds to choosing different coefficients for the
last and next--to last terms in eq (\ref{lagheavy}). Finally the 
$\tilde{\alpha}$ term corresponds to the leading term
obtained by using different values of $\alpha$ in
eqs (\ref{covderp}) and (\ref{covderq}).
Note that $(M-M^{\ast})$, $(d-d')$, $(c-c')$ and 
$\tilde{\alpha}$ all behave as $1/M$.

In addition to the terms in eq (\ref{heavy next}), which manifestly
break the heavy quark symmetry, there are, in fact, ``hidden'' 
violation terms contained in eq (\ref{lagheavy}). The explicit 
expression for the hidden terms in the model without light vectors is 
given in eq (11) of ref \cite{Ha97}. These were shown 
to exist (for the model without light vectors) in ref \cite{Ha97} 
and arise from performing a detailed $1/M$ expansion of the 
relativistic Lagrangian. In the sense of the chiral expansion these 
terms carry two derivatives, but nevertheless turn out to be 
very important numerically for the case considered. In the previous
section the numerical study has confirmed that this is also true 
when light vector mesons are included. 
It was shown ({\it cf.} fig 2 of 
ref~\cite{Ha97}) that the dependence on $d$ of the hyperfine 
splitting computed from these hidden terms using the perturbative 
approach generally matched the exact numerical calculation. Hence we 
shall not explicitly isolate the extra hidden terms due to the 
addition of the light vectors but shall content ourselves with the 
numerical treatment given in the preceding section.

In the perturbative approach the collective Lagrangian involving the
variable $A(t)$ is obtained by substituting
\begin{equation}
\overline{H}(\mbox{\boldmath $x$},t) 
= A(t) \overline{H_{\rm c}}(\mbox{\boldmath$x$})
\ ,
\label{cranked H}
\end{equation}
where 
$\overline{H_{\rm c}}(\mbox{\boldmath$x$})$ is the heavy meson 
bound--state wave function, into the heavy field Lagrangian.
Clearly this is the analog of the replacement (\ref{rotheavy}).
(The treatment of the chiral Lagrangian of the light pseudoscalars 
and vectors is the same as in section 3.) The bound--state 
wave--function is conveniently presented in the rest frame, 
$V_\mu=(1,\mbox{\boldmath $0$})$, where
\begin{equation}
\overline{H_{\rm c}} \rightarrow
\left(
\begin{array}{cc}
0 & 0 \\
\overline{h}^a_{lh} & 0 \\
\end{array}
\right)
\ ,
\end{equation}
with $a$, $l$, $h$ representing respectively the isospin, light spin
and heavy spin bivalent indices. Due to the hedgehog structure of the
soliton profiles, the calculation is simplified if we deal with a 
reduced wave--function obtained after removing the factor 
$\hat{\mbox{\boldmath$r$}}\cdot{\mbox{\boldmath $\tau$}}$:
\begin{equation}
\overline{h}^a_{lh} = \frac{u(r)}{\sqrt{M}}
\left( \hat{\mbox{\boldmath$r$}}
\cdot{\mbox{\boldmath $\tau$}}\right)_{ad}
\psi_{dl,h}
\ ,
\label{ansatz for h}
\end{equation}
where\footnote{We have removed a factor of $1/\sqrt{4\pi}$ compared 
to ref \cite{Ha97}, since it is now carried by the spherical harmonic in 
eq~(\ref{wave function:general}).
} 
$u(r)$ is a 
radial wave function, assumed to be very sharply peaked near 
$r=0$ for large $M$. In the leading order of $1/M$ there is no 
violation of the heavy quark symmetry and we may perform a 
partial wave analysis of $\psi_{dl,h}$ 
\begin{equation}
\psi_{dl,h} (g,g_3,r,k) = 
\sum_{r_3,k_3} C^{r,k;g}_{r_3,k_3;g_3}
Y^{r_3}_r \xi_{dl}(k,k_3) \chi_h
\ .
\label{wave function:general}
\end{equation}
Here $Y^{r_3}_r$ stands for the standard spherical harmonics
representing orbital angular momentum $r$ and $C$ denotes 
ordinary Clebsch--Gordan coefficients. The heavy spinor $\chi_h$ is 
trivially factored in this expression as a manifestation of the 
heavy quark symmetry. Furthermore $\xi_{dl}(k,k_3)$ represents a 
wave--function in which the light spin and isospin are added 
vectorially to give 
$\mbox{\boldmath$K$} = 
\mbox{\boldmath$I$}_{\rm light} + \mbox{\boldmath$S$}_{\rm light}$ 
with eigenvalues $\mbox{\boldmath$K$}^2=k(k+1)$. The total 
``light grand spin''
\begin{equation}
\mbox{\boldmath$g$} = \mbox{\boldmath$r$} + \mbox{\boldmath$K$}
\end{equation}
is the significant quantity in the heavy limit.
The dynamics of the model dictates that the bound-states occur for
$k=0$, in which case $\xi_{dl}(0,0) = \epsilon_{dl}/\sqrt{2}$.
The bound-state wave--function simply is
\begin{equation}
\psi_{dl,h}(0,0,0,0) = \frac{1}{\sqrt{8\pi}} \epsilon_{dl} 
\chi_{h}
\label{psi:bound}
\ .
\end{equation}
The $k=1$ unbound wave--function with no orbital excitation ($r=0$) is
\begin{equation}
\psi_{dl,h}(1,g_3,0,1) = \frac{1}{\sqrt{4\pi}} 
\xi_{dl}(1,g_3) \chi_{h}
\label{psi:unbound}
\ .
\end{equation}
When violations of the heavy quark symmetry are included,
$g$ is no longer a good quantum number.
We define the grand spin, which is a good quantum number, as,
\begin{equation}
\mbox{\boldmath$G'$} = \mbox{\boldmath$g$} + 
\mbox{\boldmath$S$}_{\rm heavy}
\ .
\end{equation}
In the notation of eqs (\ref{psi:bound}) and (\ref{psi:unbound})
we have the grand spin eigenstates
\begin{eqnarray}
\psi^{(1)}_{dl,h} (G'=G'_3=1/2)
&=& \frac{1}{\sqrt{8\pi}} \epsilon_{dl} \delta_{2h}
\ ,
\label{psi 1}\\
\psi^{(2)}_{dl,h} (G'=G'_3=1/2)
&=&
\frac{1}{\sqrt{4\pi}}
\left[
  \sqrt{\frac{2}{3}} \delta_{d1} \delta_{l1} \delta_{h1} + 
  \frac{1}{\sqrt{6}} \left(
    \delta_{d2} \delta_{l1} + \delta_{d1} \delta_{l2}
  \right) \delta_{h2}
\right]
\ .
\label{psi 2}
\end{eqnarray}
Note that in eq (\ref{psi 1}) the $G'_3=+1/2$ wave function is
$\delta_{2h}$ since the index $2$ corresponds to $+1/2$ for the
anti--quark wave--function. The two states (\ref{psi 1}) and 
(\ref{psi 2}) differ with respect to their 
$\mbox{\boldmath $g$}$ and $\mbox{\boldmath $K$}$ labels.

Now let us consider the potential for the bound-state wave--function 
in the presence of the first heavy quark symmetry violating term in 
eq (\ref{heavy next}). Substituting the $G^\prime$--eigenstates 
$\psi^{(1)}$ and $\psi^{(2)}$ from eqs (\ref{psi 1}) and 
(\ref{psi 2}) into eq (\ref{laghl}) and the first term of eq 
(\ref{heavy next}) yields, after a spatial integration, the 
potential matrix in the $\psi^{(1)}$--$\psi^{(2)}$ space:
\begin{equation}
V = - \frac{d\,F'(0)}{2}
\left( \begin{array}{cc}
3 & 0 \\ 0 & -1 
\end{array} \right)
+ \frac{M-M^{\ast}}{4}
\left( \begin{array}{cc}
0 & \sqrt{3} \\ \sqrt{3} & 2
\end{array} \right)
\ ,
\label{potential}
\end{equation}
where $F(r)$ is defined in eq (\ref{lightan}) and $F' = dF/dr$.
The first matrix shows that $\psi^{(1)}$ is bound while $\psi^{(2)}$
is unbound in the heavy spin limit. Since the second matrix gives 
mixing between $\psi^{(1)}$ and $\psi^{(2)}$ the latter must be 
included in the presence of effects which break the heavy quark 
symmetry. The diagonalized bound wave function is seen to be
\begin{equation}
\psi^{(1)} - \frac{\sqrt{3}}{8} \frac{M-M^{\ast}}{d\,F'(0)} \psi^{(2)}
\ .
\label{wave mixing}
\end{equation}
This is the proper wave--function to be ``cranked'' in order to
generate the heavy spin violation. Using it in eq (\ref{cranked H}), 
which is then substituted into the $\alpha=0$ limit of the first term 
of eq (\ref{laghl}), contributes a term in the collective Lagrangian
\begin{equation}
\frac{\chi}{2} \Omega_3 \qquad {\rm where} \qquad
\chi = \frac{M^{\ast}-M}{4d\,F'(0)} \ .
\label{chi:1st}
\end{equation}
By using the Wigner--Eckart theorem we may express this for states of
either $G'_3$ as the matrix element of the operator 
$\chi \mbox{\boldmath$\Omega$} \cdot \mbox{\boldmath$G'$}$.
For convenience we have chosen to consider our wave--function as 
representing the conjugate particle in eq (\ref{ansatz for h}). Hence 
the matrix element of $\mbox{\boldmath$G'$}$ in this section differs 
by a minus sign from that of $\mbox{\boldmath$G$}$ defined in 
eq (\ref{gspin1}).
The latter is the appropriate one when we form the total heavy baryon
spin $\mbox{\boldmath$J$} = \mbox{\boldmath$G$} +
\mbox{\boldmath$J$}^{\rm sol}$, with $J_i^{\rm sol} \equiv 
\left(\partial L/ \partial \Omega_i \right)$.
Then the collective Lagrangian, $L_{\rm coll}$ may be written
(see section 3)
\begin{equation}
L_{\rm coll} = \frac{1}{2} \alpha^2 \mbox{\boldmath$\Omega$}^2
- \chi \mbox{\boldmath$\Omega$} \cdot \mbox{\boldmath$G$}
\label{collective Lag}
\end{equation}
which again leads to the Hamiltonian (\ref{collham}).
Substituting $\alpha^2 = (3/2) \left[ m(\Delta) - m(N) \right]$ 
in eq (\ref{collham}) we get the well known formula,
{\it cf.} eq (\ref{eq:1.1})
\begin{equation}
m(\Sigma_Q^{\ast}) - m(\Sigma_Q) =
\left[ m(\Delta) - m(N) \right] \, \chi
\ .
\label{formula for splitting}
\end{equation}
The purpose in deriving this again was to explain 
the perturbative method and our notation.

Next we shall give some new perturbative ``manifest'' contributions 
to $\chi$ from eq (\ref{heavy next}).
When all these terms are included the potential $V$ in
eq (\ref{potential}) is modified so that the properly diagonalized
wave--function which replaces eq (\ref{wave mixing}) becomes
\begin{equation}
\psi^{(1)} + \epsilon \psi^{(2)} \ ,
\label{wave mix:2}
\end{equation}
with
\begin{equation}
\epsilon = 
\frac{ 
  - \frac{\sqrt{3}}{4} \left( M - M^{\ast} \right)
  + \frac{\sqrt{3}}{4} \left( d - d' \right) \, F'(0)
  + \sqrt{3} \tilde{\alpha} \omega(0)
  - \sqrt{6} \frac{c-c'}{m_V} \frac{G''(0)}{g}
}{
  2 d\, F'(0) + \frac{ 2\sqrt{2} c G''(0)}{g m_V }
}
\ .
\end{equation}
There are two types of contribution to $\chi$. The first type is 
analogous to eq (\ref{chi:1st}) and arises when 
eq (\ref{wave mix:2}) is cranked and substituted into 
eq (\ref{laghl}). The second type is obtained by 
substituting the leading order wave function $\psi^{(1)}$ into the 
$(c-c')$ and $\tilde{\alpha}$ terms in eq (\ref{heavy next}).
The complete expression for $\chi$ resulting from the ``manifest''
heavy spin violation is
\begin{eqnarray}
\lefteqn{
\chi = \epsilon 
\left[
  \frac{2}{\sqrt{3}} \left( 1 - \frac{4}{3}\alpha \right) +
  \frac{2}{3\sqrt{3}} \alpha g \left( \xi_1(0) - \xi_2(0) \right)
  - 8 \sqrt{\frac{2}{3}} \frac{c}{m_V} \varphi''(0)
\right]
}
\nonumber\\
&&
{}+
\frac{\tilde{\alpha}}{3}
\left[ 8 - 2 g \left( \xi_1(0) - \xi_2(0) \right) \right]
- 4\sqrt{2} \frac{c-c'}{m_V} \varphi''(0)
\ .
\label{manifest chi}
\end{eqnarray}
The quantities $\xi_1(0)$, $\xi_2(0)$ and
$\varphi''(0)$ are defined in eq (\ref{induced}).
This formula may be useful for quickly estimating the effects of
heavy spin violation in the coupling constants, which were not
explicitly given in the previous discussion. Unfortunately there 
is no determination of the magnitude of these effects from the 
mesonic sector at present. In ref \cite{Ha97} the discussion of the 
``hidden'' heavy contributions to $\chi$ was given for the 
Lagrangian with only light pseudoscalars. 

The hyperfine splitting just discussed is for the ground state or
P--wave heavy baryons. It is of some interest to briefly consider 
the negative parity heavy baryons with one unit of orbital 
excitation.  In the heavy spin limit these bound states 
correspond to the $r=1$ and $k=0$ choice in 
eq (\ref{wave function:general}):
\begin{equation}
\psi_{dl,h}(1,g_3,1,0) = 
\frac{\epsilon_{dl}}{\sqrt{2}} Y_1^{g_3} \chi_h \ .
\end{equation}
The spin, $\mbox{\boldmath$J$}_{\rm light}$ of the ``light cloud''
part of the heavy baryon is gotten by adding this $g=1$ piece to the 
soliton spin $\mbox{\boldmath$J$}^{\rm sol}$. For the $I=0$ (which 
implies $J^{\rm sol}=0$) heavy baryons one finds $J_{\rm light}=1$ 
and the degenerate multiplet
\begin{equation}
\left\{ \Lambda_Q'(1/2) \,,\, \Lambda_Q'(3/2) \right\}
\ .
\label{Lambda prime}
\end{equation}
For the $I=J^{\rm sol}=1$ heavy baryons, $J_{\rm light}$ can be 
either $0$, $1$ or $2$ and we find the degenerate heavy multiplets
\begin{eqnarray}
&& \qquad \Sigma_Q'(1/2) \ ,
\nonumber\\
&& \left\{ \Sigma_Q'(1/2) \,,\, \Sigma_Q'(3/2) \right\} \ ,
\nonumber\\
&& \left\{ \Sigma_Q'(3/2) \,,\, \Sigma_Q'(5/2) \right\}
\ .
\label{excited Sigmas}
\end{eqnarray}
In general, the situation is even more complicated and further
discussion will be given elsewhere.  At present there are 
experimental candidates\cite{PDG96} for a negative parity spin 
$1/2$ baryon $\Lambda_c'$ at $2593.6\pm1.0$\,MeV and a negative 
parity spin $3/2$ baryon $\Lambda_c'$ at $2626.4\pm0.9$\,MeV.

Since experimental information is available, it is especially
interesting to consider the splitting between the two $\Lambda_Q'$ 
states in eq (\ref{Lambda prime}). This splitting stems from the 
violation of the heavy quark symmetry. For the $\Lambda_Q$ type 
states the total spin coincides with the grand spin 
$\mbox{\boldmath$G$}$ so that eq (\ref{Lambda prime}) may be
alternatively considered a $G=1/2$, $G=3/2$ multiplet. Since the 
good quantum number is $G$, we may in general expect the hyperfine
parameter $\chi$ to depend on $G$. The collective Hamiltonian takes 
the form
\begin{equation}
H_{\rm coll} = \frac{ \left(\mbox{\boldmath$J$}^{\rm sol} + 
\chi_G \mbox{\boldmath$G$} \right)^2}{ 2\alpha^2}
\ .
\label{collective Hamiltonian}
\end{equation}
On general grounds we see that for the case of the $\Lambda_Q'$'s the
collective Hamiltonian contribution to the hyperfine splitting will be
suppressed. Setting $\mbox{\boldmath$J$}^{\rm sol}=0$ in 
eq (\ref{collective Hamiltonian}) shows that the hyperfine splitting
is of order $(\chi^2)$ or equivalently of order $(1/M^2)$.
Unlike the ground state which involves only the $G=1/2$ P--wave
channel, there is another possibility for hyperfine splitting here.
It is allowed for the $G=1/2$ and $G=3/2$ bound state energies to
differ from each other. In the Lagrangian with only light 
pseudoscalars this does not happen and the 
$\Lambda_Q'(1/2) - \Lambda_Q'(3/2)$ splitting is of order
$1/M^2$. However when light vectors are added, there are ``hidden''
$1/M$ terms, which violate the heavy quark symmetry as {\it e.g.}
\begin{equation}
i\,\,
\mbox{\rm Tr} \, 
\left[
  \sigma_{\alpha\mu} H \gamma_\nu F^{\mu\nu}(\rho)
  D^\alpha \overline{H}
\right] \, + \, \mbox{\rm h.c.}
\ .
\end{equation}
This term is likely to generate splitting for the multiplet 
(\ref{Lambda prime}) to order $1/M$ by giving different binding
energies to the $G=1/2$ and $G=3/2$ channels.
It would be very interesting to investigate this in more detail.

Finally, we add a remark concerning an amusing conceptual feature in
the computation of hyperfine splitting among the five $\Sigma_Q'$'s in
eq (\ref{excited Sigmas}). The total angular momentum of each state
is given by
\begin{equation}
\mbox{\boldmath$J$} = 
\underbrace{
  \mbox{\boldmath$J$}^{\rm sol} + \mbox{\boldmath$g$} 
}_{\mbox{\boldmath$J$}_{\rm light}}
\!\!\!\!\!
\overbrace{
  \ \ \ + \mbox{\boldmath$S$}_{\rm heavy}
}^{\mbox{\boldmath$G$}}
\ ,
\label{recoupling formula}
\end{equation}
where we are now considering each operator to be acting on the
wave--function rather than its complex conjugate. We have illustrated 
two different intermediate angular momenta which can alternatively 
be used to label the final state. 
If $\mbox{\boldmath$J$}_{\rm light}$ is used, we get the
heavy-spin multiplets in eq (\ref{excited Sigmas}). 
On the other hand, when the hyperfine splitting is turned on, the 
choice $\mbox{\boldmath$G$}$ is convenient, because it remains a 
good quantum number. According to the laws of quantum mechanics, we 
cannot simultaneously use both to specify the states, since the 
commutator
\begin{equation}
\left[ 
  \, \mbox{\boldmath$J$}_{\rm light}^2 \ , \ 
  \mbox{\boldmath$G$}^2 \,
\right]
= 4 i \mbox{\boldmath$J$}_{\rm light} 
\mbox{\boldmath$\cdot$}
\left(\mbox{\boldmath$S$}_{\rm heavy}
\mbox{\boldmath$\times$} \mbox{\boldmath$g$}\right)
\label{commutator}
\end{equation}
is generally non--vanishing. This means that we cannot uniquely 
trace the splitting of, say, 
the $\left\{\Sigma^\prime_Q(1/2),\Sigma^\prime_Q(3/2)\right\}$
heavy multiplet in eq~(\ref{excited Sigmas}), as hyperfine
splitting interactions are turned on. Physically, this causes 
a mixing between the $\Sigma^\prime_Q$'s of the same spin. Rather,
we must look at the whole pattern of the five masses. On the other hand,
the problem simplifies for the computation of the ground state
hyperfine splitting in eq (\ref{formula for splitting}).
In that case the bound state wave function is characterized by 
$\mbox{\boldmath$g$}=0$. Thus the commutator in 
eq (\ref{commutator}) vanishes, and it is ``trivially'' possible to 
track the hyperfine splitting as a mass difference.

\bigskip
\stepcounter{chapter}
\leftline{\Large \it 6. Discussion and Conclusions}
\smallskip

In the framework of the bound state approach we have studied 
the hyperfine splitting for baryons containing a heavy quark.
In this approach a heavy baryon is constructed from a heavy meson
configuration bound in the background of a (chiral) soliton. Here we 
have limited ourselves to heavy mesons in the S-- and P-- 
wave channels, which exhibit the strongest binding. The study has
been motivated by the earlier observation that light vector meson 
fields are required in the soliton configuration in order to 
reasonably describe the spectra of both the light and the heavy 
baryons when all available information on coupling constants of 
the elementary mesons is incorporated. The inclusion of light vector 
mesons causes some technical difficulties because field components 
which vanish classically are induced when time--dependent
collective coordinates are introduced in order to generate states 
with good spin and isospin from the soliton. One might argue that 
the better agreement in the vector meson model is due to an 
additional parameter $\alpha$; however, we have observed that the 
agreement is achieved for quite a wide range of this parameter. 
In fact the binding energies vary by only about $100{\rm MeV}$ 
in the range $-0.1\le\alpha\le0.3$. On the other hand the 
discrepancy between the empirical data and the predictions obtained 
from the Skyrme model soliton is about twice as large. In addition 
the vector meson model reasonably reproduces the relative (to the 
nucleon) masses for both the charm and the bottom sector. Furthermore 
the mass difference within a given heavy multiplet, {\it i.e.} the 
hyperfine splitting, has turned out not to be very sensitive to 
that parameter either.

We also have estimated kinematical corrections by substituting 
the reduced masses. The comparison with the empirical data has 
certainly indicated that these corrections are important.
This simple non-relativistic substitution fails, however, to 
satisfy the heavy quark limit result, which states that the 
hyperfine splitting should vanish for infinitely large quark masses.
It thus seems interesting to further explore the kinematical 
corrections.

As an extension of earlier work \cite{Ha97} we have illuminated the 
systematics of the $1/M$ expansion of the hyperfine splitting. The 
main conclusion is that the dominant contribution stems from terms in 
the relativistic Lagrangian which do not manifestly break the 
heavy quark symmetry. 

We have furthermore observed that the heavy quark approach does
not seem to be suitable for the strange sector. The binding energies
simply turned out too low for reasonable predictions of the mass 
differences between the heavy baryons and the nucleon.

On the other hand an interesting path to pursue would be the 
extension of the light sector to flavor $SU(3)$. This would make 
possible the description of baryons like $\Xi_c$ or $\Xi^*_b$. 
This would in particular be interesting for the issue of 
flavor symmetry breaking \cite{Mo94}. Unfortunately the vector 
meson model for three flavors requires the introduction of 
additional induced components for the strange degrees of 
freedom \cite{Pa91}.

\bigskip
\leftline{\Large\it Acknowledgements}
\smallskip

This work was supported in part by the US DOE contract number
DE--FG--02--85ER 40231 and by the Deutsche Forschungsgemeinschaft 
(DFG) under contract Re 856/2--2.

\bigskip
\appendix
\stepcounter{chapter}
\leftline{\Large \it Appendix A: Bound State Lagrangian}
\smallskip

In this appendix we present the Lagrangian for the
{\it ans\"atze} (\ref{pansatz})--(\ref{sansatz2}) of the bound
heavy mesons. These expressions have already been presented in 
appendix A of ref \cite{Sch96}. Unfortunately some typographical 
errors have occurred in the formulas reported there. It is therefore 
appropriate to list the corrected expressions. The present notation 
corresponds to eq (\ref{laggen}).

Substituting (\ref{sansatz}) and (\ref{sansatz2}) 
in (\ref{lagheavy}) gives for the
S--wave channel
\begin{eqnarray}
I_\epsilon^{(S)}&=&\int dr r^2 \Bigg(\Phi^{\prime2}+
\left[M^2-\left(\epsilon-\frac{\alpha}{2}\omega\right)^2
+\frac{R_\alpha^2}{2r^2}\right]\Phi^2
+M^{*2}\left[\Psi_1^2+2\Psi_2^2-\Psi_0^2\right]
\nonumber \\ &&
+\frac{2}{r^2}\left[r\Psi_2^\prime+\Psi_2
-\frac{R_\alpha+2}{2}\Psi_1\right]^2+\frac{R_\alpha^2}{r^2}\Psi_2^2
-\left[\Psi_0^\prime
-\left(\epsilon-\frac{\alpha}{2}\omega\right)\Psi_1\right]^2
\nonumber \\ &&
-2\left[\left(\epsilon-\frac{\alpha}{2}\omega\right)\Psi_2
-\frac{R_\alpha+2}{2r}\Psi_0\right]^2
+2Md\left[F^\prime\Psi_1+\frac{2}{r}{\rm sin}F \Psi_2\right]\Phi
\nonumber \\ &&
+2d\Bigg\{F^\prime\left[\frac{1}{r}
\left(1+{\rm cos}F\right)\Psi_0\Psi_2
-\left(\epsilon-\frac{\alpha}{2}\omega\right)\Psi_2^2\right]
\nonumber \\ && \hspace{1cm}
+\frac{2}{r}{\rm sin}F\left[\Psi_2\Psi_0^\prime
-\left(\epsilon-\frac{\alpha}{2}\omega\right)\Psi_1\Psi_2
+\frac{R_\alpha+2}{2r}\Psi_0\Psi_1\right]\Bigg\}
\nonumber \\ &&
+\frac{4\sqrt{2}cM}{gm_V}\left[\omega^\prime\Psi_0\Psi_1
+\frac{2G^\prime}{r}\Psi_1\Psi_2
+\frac{G}{r^2}\left(G+2\right)\Psi_2^2\right]
\nonumber \\ &&
-\frac{4\sqrt{2}c}{gm_V}
\Bigg\{\frac{\omega^\prime}{r}R_\alpha\Phi\Psi_2
+\frac{G}{r^2}\left(G+2\right)\Psi_0\Phi^\prime
+\frac{G^\prime}{r}R_\alpha\Phi\Psi_0
\nonumber \\ &&  \hspace{2cm}
+\left(\epsilon-\frac{\alpha}{2}\omega\right)
\left[\frac{2G^\prime}{r}\Psi_2
+\frac{G}{r^2}\left(G+2\right)\Psi_1\right]\Phi\Bigg\}
\Bigg).
\label{swlag}
\end{eqnarray}
Here a prime indicates a derivative with respect to the radial
coordinate $r$. Furthermore the abbreviation
$R_\alpha={\rm cos}F-1+\alpha\left(1+G-{\rm cos}F\right)$
has again been used. 

For the P--wave channel one obtains upon substitution of the
{\it ansatz} (\ref{pansatz}) and (\ref{pansatz2})
\begin{eqnarray}
I_\epsilon^{(P)}&=&\int dr r^2 \Bigg(\Phi^{\prime2}+
\left[M^2-\left(\epsilon-\frac{\alpha}{2}\omega\right)^2
+\frac{2}{r^2}\left(1+\frac{1}{2}R_\alpha\right)^2\right]\Phi^2
+M^{*2}\left[\Psi_1^2+\frac{1}{2}\Psi_2^2-\Psi_0^2\right]
\nonumber \\ &&
+\frac{1}{2}\left[\Psi_2^\prime-\frac{1}{r}\Psi_2\right]^2
+\frac{1}{r}R_\alpha\Psi_1\Psi_2^\prime
+\frac{1}{r^2}R_\alpha\left(\Psi_1+\Psi_2\right)\Psi_2
+\frac{1}{2r^2}R_\alpha^2\left(\Psi_1^2+\frac{1}{2}\Psi_2^2\right)
\nonumber \\ &&
-\left[\Psi_0^\prime-
\left(\epsilon-\frac{\alpha}{2}\omega\right)\Psi_1\right]^2
-\frac{1}{2}\left[\frac{R_\alpha}{r}\Psi_0
+\left(\epsilon-\frac{\alpha}{2}\omega\right)\Psi_2\right]^2
\nonumber \\ &&
-d\Bigg\{\frac{2}{r}{\rm sin}F\left[\Psi_2\Psi_0^\prime
-\frac{R_\alpha}{r}\Psi_0\Psi_1
-\left(\epsilon-\frac{\alpha}{2}\omega\right)\Psi_1\Psi_2\right]
\nonumber \\ && \hspace{1cm}
+\frac{F^\prime}{r}\left[\frac{r}{2}
\left(\epsilon-\frac{\alpha}{2}\omega\right)\Psi_2^2
-\left(1-{\rm cos}F\right)\Psi_0\Psi_2\right]\Bigg\}
+2Md\left[F^\prime\Psi_1-\frac{{\rm sin}F}{r}\Psi_2\right]\Phi
\nonumber \\ &&
+\frac{2\sqrt{2}cM}{gm_V}\left[2\omega^\prime\Psi_0\Psi_1
-\frac{2G^\prime}{r}\Psi_1\Psi_2
+\frac{G}{2r^2}\left(G+2\right)\Psi_2^2\right]
\nonumber \\ &&
-\frac{4\sqrt{2}c}{gm_V}\Bigg\{
\frac{1}{r^2}\left(\epsilon-\frac{\alpha}{2}\omega\right)
\left[G\left(G+2\right)\Psi_1-rG^\prime\Psi_2\right]\Phi
-\frac{\omega^\prime}{r}\left[1+\frac{R_\alpha}{2}\right]\Psi_2
\nonumber \\ && \hspace{2cm}
+\frac{1}{r^2}\left[G\left(G+2\right)\Phi^\prime
+G^\prime\left(2+R_\alpha\right)\Phi\right]\Psi_0\Bigg\}
\Bigg).
\label{pwlag}
\end{eqnarray}
The typographical errors in ref \cite{Sch96} only affect the 
expressions involving the parameter $c$.

\bigskip
\stepcounter{chapter}
\leftline{\Large \it Appendix B: Hyperfine Parameters}
\smallskip

In this appendix we give the explicit expressions for the 
hyperfine splitting parameters used in section 4. 
For convenience we employ additional abbreviations with 
regard to the light meson profiles defined in 
eqs (\ref{lightan}) and (\ref{induced})
\be
V_1&=&{\rm cos}F-\alpha\left(\xi_1-1+{\rm cos}F\right)\ ,
\nonumber \\
V_2&=&1-\alpha\left(\xi_1+\xi_2\right) \ .
\nonumber
\ee
The explicit expression for the P--wave hyperfine parameter,
which enters the mass formula for the even parity heavy 
baryon (\ref{massp}), reads
\be
\chi_P&=&\frac{2}{3}\int_0^\infty dr\ r^2\ \rho_{\chi}^{(P)}(r)
\label{eq3} \\
\rho_{\chi}^{(P)}(r)\hspace{-5pt}&=& \hspace{-4pt}
\left[\eal \left(V_2-2V_1\right)
-\frac{2\alpha}{r^2}\left(2+R_\alpha\right)\varphi\right]\Phi^2
\nonumber \\ && \hspace{-4pt}
+\left(2V_1+V_2\right)\left[\eal\Psi_1-\Psi_0^\prime\right]\Psi_1
\nonumber \\ && \hspace{-4pt}
-\frac{1}{2}\left(V_2\Psi_2+\frac{4\alpha}{r}\varphi\Psi_0\right)
\left[\eal\Psi_2+\frac{R_\alpha}{r}\Psi_0\right]
\nonumber \\ && \hspace{-4pt}
+\frac{2\alpha}{r}\varphi\Psi_1
\left(\Psi_2^\prime+\frac{1}{r}\Psi_2+\frac{R_\alpha}{r}\Psi_1\right)
-\frac{\alpha}{r^2}\left(2+R_\alpha\right)\varphi\Psi_2^2
+4Md\ {\rm sin}F\Phi\Psi_0
\nonumber \\ && \hspace{-4pt}
-\frac{d}{r}
\left\{{\rm sin}F\left[\left(2+R_\alpha+V_1\right)\Psi_1\Psi_2
-\frac{4\alpha}{r}\varphi\Psi_0\Psi_1\right]
+F^\prime\left[\frac{r}{4}V_2\Psi_2^2+
2\alpha\varphi\Psi_0\Psi_2\right]\right\}
\nonumber \\ && \hspace{-4pt}
-\frac{4\sqrt2cM}{gm_\rho}\left\{
\left(3\xi_1^\prime+\xi_2^\prime\right)\Psi_0\Psi_1
+\frac{G}{r}\left(2-2\xi_1-\xi_2\right)\Psi_0\Psi_2
+\frac{2}{r}\varphi^\prime\Psi_1\Psi_2
+\frac{1}{r^2}\varphi\Psi_2^2\right\}
\nonumber \\ && \hspace{-4pt}
-\frac{4\sqrt2c}{gm_\rho}
\Bigg\{\eal\left(\frac{4}{r^2}\varphi\Psi_1+
\frac{2}{r}\varphi^\prime\Psi_2\right)\Phi
+\left(V_1-\frac{V_2}{2}\right)
\left[\frac{G}{r^2}\left(G+2\right)\Psi_1
-\frac{G^\prime}{r}\Psi_2\right]\Phi
\nonumber \\ && \hspace{-4pt}
+\frac{G}{r^2}\left(2+R_\alpha\right)
\left(2\xi_1+\xi_2-2\right)\Phi\Psi_1
+\frac{2}{r^2}\left[2\varphi\Phi^\prime
+\left(2+R_\alpha\right)\varphi^\prime\Phi
-\alpha G^\prime\varphi\Phi\right]\Psi_0
\nonumber \\ && \hspace{-4pt}
-\frac{1}{r}\left[\left(G+2\right)\xi_2\Phi^\prime
+\left(1+\frac{1}{2}R_\alpha\right)
\left(\xi_1^\prime+\xi_2^\prime\right)\Phi
-\alpha\omega^\prime\varphi\Phi\right]\Psi_2\Bigg\}
\ee
\bigskip
It can easily be verified that the terms involving
$\eal\approx M$ cancel when the heavy limit relations 
for the radial functions (\ref{pwhl}) are substituted.
Taking into account that the radial functions 
which parametrize the heavy meson wave--functions are 
normalized to $1/\sqrt{|\epsilon|}\approx 1/\sqrt{M}$ 
({\it cf.} eq (\ref{normp})), it is obvious that the hyperfine 
parameter $\chi_P$ vanishes in the heavy quark limit..

The hyperfine parameter for the odd parity baryon 
({\it cf.} eq (\ref{masss})) is found to be
\be
\chi_S&=&\frac{2}{3}\int_0^\infty dr\ r^2\ \rho_{\chi}^{(S)}(r)
\label{eq5} \\
\rho_{\chi}^{(S)}(r)\hspace{-5pt}&=& \hspace{-4pt}
\left[\eal(2V_1+V_2)+\frac{2\alpha}{r^2}R_\alpha\varphi\right]\Phi^2
+(2V_1-V_2)\left[\Psi_0^\prime-\eal\Psi_1\right]\Psi_1
\nonumber \\ && \hspace{-4pt}
+2\left[\frac{R_\alpha+2}{2r}\Psi_0-\eal\Psi_2\right]
\left(V_2\Psi_2+2\frac{\alpha}{r}\varphi\Psi_0\right)
+4\frac{\alpha}{r^2}R_\alpha\varphi\Psi_2^2
\nonumber \\ && \hspace{-4pt}
+4\frac{\alpha}{r^2}\varphi
\left[r\Psi_2^\prime+\Psi_2-\frac{R_\alpha+2}{2}\Psi_1\right]\Psi_1
-4Md\ {\rm sin}F\Phi\Psi_0
\nonumber \\ && \hspace{-4pt}
-d\left\{
F^\prime\left(V_2\Psi_2+\frac{4\alpha}{r}\varphi\Psi_0\right)\Psi_2
+\frac{2}{r}{\rm sin}F\left[\left(V_1+R_\alpha\right)\Psi_2
+\frac{2\alpha}{r}\varphi\Psi_0\right]\Psi_1\right\}
\nonumber \\ && \hspace{-4pt}
-\frac{4\sqrt2cM}{m_V g}\left\{
\left(\xi_2^\prime-\xi_1^\prime\right)\Psi_0\Psi_1
+\frac{2}{r}\left(G+2\right)\xi_2\Psi_0\Psi_2
+\frac{4}{r}\varphi^\prime\Psi_1\Psi_2
+\frac{4}{r^2}\varphi\Psi_2^2\right\}
\nonumber \\ && \hspace{-4pt}
-\frac{4\sqrt2c}{m_V g}\Bigg\{
\frac{2}{r}G\left(2\xi_1+\xi_2-2\right)\Phi^\prime\Psi_2
+\frac{R_\alpha}{r^2}\left(G+2\right)\xi_2\Phi\Psi_1
\nonumber \\ && \hskip1cm
+\frac{1}{r}\left[R_\alpha\left(\xi_1^\prime+\xi_2^\prime\right)
+2\alpha\omega^\prime\varphi\right]\Phi\Psi_2
+\frac{2}{r^2}\left[2\varphi\Phi^\prime
-\left(R_\alpha\varphi^\prime
-\alpha G^\prime\varphi\right)\Phi\right]\Psi_0
\nonumber \\ && \hskip1cm
+\frac{4}{r^2}\eal\left(\varphi\Psi_1+r\varphi^\prime\Psi_2\right)\Phi
\nonumber \\ && \hskip1cm
-\left(V_1+\frac{1}{2}V_2\right)
\left[\frac{1}{r^2}G\left(G+2\right)\Psi_1
+\frac{2}{r}G^\prime\Psi_2\right]\Phi\Bigg\}\ .
\ee
Again, it can easily be verified that the terms involving
$\eal\approx M$ vanish when the heavy limit relations 
for the radial functions (\ref{swhl}) are substituted.
With regard to the normalization condition (\ref{norms}) 
the S--wave hyperfine parameter also behaves like
$\chi_S\sim 1/M$ in the heavy quark limit.

\end{document}